\documentclass{article}

\usepackage{graphicx,natbib,amsmath,amsfonts,amsthm,bm}
\usepackage[dvipsnames]{xcolor}
\usepackage{soul}

\newtheorem{theorem}{Theorem}

\newcommand{\argmin}{\ensuremath{\operatornamewithlimits{argmin}}}

\begin{document}

\title{Unrestricted Permutation forces Extrapolation: \\ Variable Importance Requires at least One More Model \\  or \\ There Is No Free Variable Importance}

\author{Giles Hooker\footnote{University of California, Berkeley, ghooker@berkeley.edu}, Lucas Mentch\footnote{University of Pittsburgh, lkm31@pitt.edu}, and Siyu Zhou\footnote{University of Pittsburgh, siz25@pitt.edu}}
\date{}
\maketitle

%
%
%

\begin{abstract}
This paper reviews and advocates against the use of permute-and-predict (PaP) methods for interpreting black box functions.  Methods such as the variable importance measures proposed for random forests, partial dependence plots, and individual conditional expectation plots remain popular because they are both model-agnostic and depend only on the pre-trained model output, making them computationally efficient and widely available in software.  However, numerous studies have found that these tools can produce diagnostics that are highly misleading, particularly when there is strong dependence among features.  The purpose of our work here is to (i) review this growing body of literature, (ii) provide further demonstrations of these drawbacks along with a detailed explanation as to why they occur, and (iii) advocate for alternative measures that involve additional modeling.  In particular, we describe how breaking dependencies between features in hold-out data places undue emphasis on sparse regions of the feature space by forcing the original model to extrapolate to regions where there is little to no data.  We explore these effects across various model setups and find support for previous claims in the literature that PaP metrics can vastly over-emphasize correlated features in both variable importance measures and partial dependence plots.  As an alternative, we discuss and recommend more direct approaches that involve measuring the change in model performance after muting the effects of the features under investigation.
\end{abstract}


\section{Introduction}
Machine learning methods have proved to be enormously successful tools for making predictions from data. However, most of these methods produce algebraically complex models that provide little or no insight into how they arrive at their predictions. As a consequence, many researchers have suggested methods to ``X-ray the black box'' in order to provide insight into which input features are important and how they effect predictions; a task so precarious that others have advocated against the practice altogether \citep{rudin2019stop}.

This paper reviews methods based on permuting feature values or otherwise investigating the effect that changing the values of a feature has on predictions. Our message can be simply summarized by
\begin{quotation}
{\em When features in the training set exhibit statistical dependence, permute-and-predict methods can be highly misleading when applied to the original model.}
\end{quotation}

Permutation methods are some of the oldest, most popular, and computationally convenient means of understanding complex learning algorithms.  In this paper, we will focus primarily on three commonly-used techniques:
\begin{description}
\item[Variable Importance]  These methods are designed to provide a score for each feature based on how much difference replacing the feature with noise makes in predictive accuracy. In particular, \cite{randomforests} introduced the idea of measuring the importance of the $j$th feature by permuting its values in the training data and examining the corresponding drop in predictive accuracy when these new data are used in a model built with the original training data.  

Given a training set consisting of a matrix of feature values $X$ with rows $\bm{x}_i$ giving each observation and corresponding response vector $\bm{y}$, let $X^{\pi,j}$ be a matrix achieved by randomly permuting the $j$th column of $X$.  Using $L(y_i,f(\bm{x}_i))$ as the loss for predicting $y_i$ from $f(\bm{x}_i)$, the importance of the $j$th feature can be defined as
    \[
    \mbox{VI}_j^{\pi} = \sum_{i=1}^N L(y_i, f(\bm{x}^{\pi,j}_i)) - L(y_i,f(\bm{x}_i))
    \]
so as to measure the increase in loss due to replacing $x_{ij}$ with a value randomly chosen from the (marginal) distribution of feature $j$.

\cite{randomforests} designed the method specifically for use with random forests and considered {\em out-of-bag} (OOB) loss in which variable importance is averaged over trees, measured with that part of the data not used to construct the corresponding tree.  For more general learners, either training or test loss can be used.   Since being introduced, numerous variations on this idea have appeared. As one recent notable example, \citet{fisher2019all} considered averaging over all possible permutations and derived concentration bounds based on $U$-statistics. Nonetheless, each of these methods utilizes the same general permute-and-predict (PaP) structure \citep[][also provide a version based on conditional permutatations]{fisher2019all}; our simulations employ both the $\mbox{VI}_j^{\pi}$ and OOB importances when examining random forests.

\item[Partial Dependence Plots (PDPs)] \cite{Friedman2001} suggested examining the effect of feature $j$ by plotting the average prediction as the feature is changed.  Specifically, letting $X^{x,j}$ be the matrix of feature values where the $j$th entry of every row has been replaced with value $x$, we define the partial dependence function
    \[
    \mbox{PD}_j (x) = \frac{1}{N} \sum_{i=1}^N f(\bm{x}^{x,j}_i)
    \]
    as the average prediction made with the $j$th feature replaced with the value $x$. Since these are univariate functions (multivariate versions can be defined naturally) they can be readily displayed and interpreted.

\item[Individual Conditional Expectation (ICE) Plots] \citet{Goldstein2015} suggested a refined version of partial dependence plots that involves simply plotting out the functions
    \[
    \mbox{ICE}_{i,j} (x) = f(\bm{x}^{x,j}_i)
    \]
    that correspond to tracing out the prediction given to any one example as the $j$th feature is changed. PDPs are then exactly the average of the corresponding ICE plots, but the latter allows an investigation in how the effect of feature $j$ may change for different combinations of the remaining inputs.  When $N$ is very large, a random selection of ICE plots can be presented as examples. \cite{Goldstein2015} also described how these ICE plots can potentially be used to detect the kind of extrapolation we discuss in detail in this paper.
\end{description}

These techniques are attractive for a number of reasons:  they are each computationally cheap, requiring $O(N)$ operations, apply to the $f(\bm{x})$ derived from any learning method, and have no tuning parameters.  These computational benefits make them easily programmable and as a result, they are widely available across a multitude of software implementations.  From a statistical standpoint, they rely only on averages and are thus generally very stable. Moreover, at a high level, the approach has an intuitive feel that can give practitioners a (perhaps unwarranted) sense of confidence. For these reasons, they have been frequently adopted across a wide variety of scientific domains and still appear to remain the go-to tool of choice for most applied scientists utilizing black-box models.

However, procedures like these based on the PaP structure have been shown to exhibit serious flaws, especially when significant correlations exist between features.  \citet{Strobl2007} investigate classification and note that the importance measures based on permuted OOB error in CART built trees are biased toward features that are correlated with other features and/or have many categories and further suggest that bootstrapping exaggerates these effects.  \citet{Archer2008} explore a similar set-up and also note improved performance when (true) signal features -- those actually related to the response -- are uncorrelated with noise features.  \citet{Nicodemus2010} investigates these claims of feature preference in a large scale simulation study and again find that the OOB measures overestimate the importance of correlated predictors.



In explaining the observed behavior of variable importance measures, we need to distinguish two effects:
\begin{itemize}
\item The use of a permutation-based measure of importance to summarize a known and fixed function when features are correlated. Permutation measures ignore the correlation between $x_1$ and $x_2$, treating each as being manipulatable separately from the other in ways that may not be realistic.

    Given a fixed $f(x)$ then with increasing sample size we can define the target of $\mbox{VI}_j^{\pi}$ as
\[
\mbox{VI}_j^{\pi}(f) = E_{y,\bm{x}}\left[ L(y,f_{-j}(\bm{x}_{-j})) -  L(y,f(\bm{x})) \right]
\]
where
\[
f_{-j}(\bm{x}_{-j}) = \int f(\bm{x}) p_j(x_j) dx_j
\]
and $p_j(x_j)$ is the marginal distribution of $x_j$. Similarly, partial dependence has a general target of
\[
f_{j}(x_j) = \int f(\bm{x}) p_{-j}(\bm{x}_{-j}) d\bm{x}_{-j}.
\]

\item The effect of applying permutation methods to {\em estimated} functions. Here, if we treat permutation importance applied to the underlying relationship as the quantity to be estimated, and our application of permutation importance to an estimated random forest as an estimate, we can observe considerable bias due to measuring the random forest at extrapolation.
\end{itemize}
The first of these is inescapable, and the appropriateness of any particular importance summary will depend on the application at hand; arguments such as those in \citet{Strobl2007} can be seen as focussed on this question.  In contrast, we believe the second of these is at least as important and follow \citet{Hooker2007} in assigning the effect to extrapolation. For example, suppose features $x_1$ and $x_2$ are strongly positively dependent, there will be no training examples which pair a large value of $x_1$ with a small value of $x_2$ or {\em vice versa}. Thus, the predictions made in the upper-left corner of $(x_1,x_2)$ space will mostly depend on the extrapolation behavior of the particular learning method employed. And as we demonstrate in the following sections, permutation-based methods place significant weight on exactly these predictions.  As a concrete example: an evaluation of the importance of pregnancy status in a model that also includes gender would result in the evaluation of the response of pregnant men as often as pregnant women.


Our thesis here is illustrated by considering a linear model:
\[
f(x) = \beta_0 + \sum_{j=1}^p \beta_j x_j
\]
where, if each covariate $x_j$ has variance 1, its permutation importance is given by $\beta_j^2$, regardless of the correlation among the features.  While this is by no means the only way to define importance for a linear model, it does correspond to the familiar incantation of ``the change in $y$ for one unit change in $x_j$, keeping all else fixed'' and could be construed as justifying permute-and-predict measures.  In this model, we can also regard $\beta_j^2$ (or its rank among coefficients) as the target of estimation when we initially obtain a random forest or a neural network from data generated from this model and then apply permutation importance to it. In our experiments, we observe that this estimate is biassed upwards for covariates that are correlated with each other, and that this bias increases with correlation.  The alternative measures for which we advocate imply a different notion of variable importance in linear models, but they do not extrapolate and do not suffer from the same discrepancy between importance applied to a known underlying model, and importance applied to a machine learning estimate of it.

For these reasons, we argue that these methods can be misleading. Consider a situation in which $x_1$ and $x_2$ are strongly correlated and a backwards elimination strategy is being used as a covariate selection procedure (e.g. \cite{Diaz2006} built a sequence of random forests removing the least important features at each step). Our examples provide a situation in which $x_1$ and $x_2$ would {\em both} be retained ahead of features that are ranked as more important than them in terms of the underlying linear model. Such a procedure may then result in a collection of strongly correlated features, each of which is individually only weakly predictive (in combination with the other retained features) while more strongly predictive features are discarded.

Our argument implicitly connects variable importance to notions of statistical inference; variable importances are an estimate for the quantity that would have been obtained by making the same calculation with unknown true response function. We are, as such, interested in the {\em statistical} properties of a machine learning method, as opposed to conducting a ``model audit'' in which an estimated model is considered fixed and we merely wish to summarize its behavior.  This is similar to the distinction in \citet{fisher2019all} between \emph{model reliance} and \emph{model class reliance} which provides generalized notion confidence interval for variable importance.  That said, we restrict our attention here to {\em bias due to extrapolation}; measures of variability will also depend on the machine learning method being used. Confidence intervals can be calculated in the case of ensemble methods such as random forests by relating them to $U$-statistics \citep[e.g.][]{ishwaran2019standard}. The notion of model class reliance in \citet{fisher2019all} provides a highly adaptable analogue of profile confidence intervals for methods based on optimization. These can be employed with any notion of variable importance including either the permute-and-predict methods we critique here, or using conditional permutations. While we do not investigate these measures here our results would advocate for using the latter.

Our goals in this paper are to (i) review the growing body of literature on this topic, (ii) provide an extended, more detailed examination of the effect of these diagnostic tools along with an explanation for this behavior when applied to random forests, and (iii) advocate for recently developed importance measures that avoid extrapolation, either by generating perturbed data differently, or re-learning models.  In addition to random forests, we also examine the behavior of neural networks where different, but similarly problematic, behavior is observed.

While permute-and-predict measures produced biased estimates, there are alternatives that do not involve measuring the behavior of models far from their training data, and we explore two strategies to do this. The first of these is to generate perturbed versions of the feature of interest {\em conditionally} on the vector of the remaining features, (as opposed to independently of them), thus avoiding distorting the feature distribution; the approaches of \citet{Strobl2007} fall within this class.  A more general approach is {\em re-learning}, which
generally involves measuring the drop in model performance when the effects of the features in question are muted.  The particular way in which the muting and testing occurs varies slightly between procedures appearing in the recent literature -- \cite{Mentch2016,Mentch2017} suggest permuting said features and rebuilding the model, the leave-out-covariates (LOCO) approach in \cite{Lei2018} simply drops them, and the conditional randomization test in \cite{Candes2018} substitutes those features in question with randomized replacements (knockoffs) sampled conditional on the remaining features.  In each case, however, rather than extrapolate the original model, a new  model is \emph{re-learned} using the data $(\bm{y},X^{\pi}_j)$ and the change in performance is tested for statistical significance.  In fact, all these strategies result in the same population target value if we use squared error loss, and our simulations suggest they do not exhibit the same biases as permute-and-predict. Note that either generating conditional random variables, or re-learning requires the use of an additional model and therefore computational or modeling effort, hence our subtitle.

We stress that these re-learning procedures, while a substantial improvement, are themselves not entirely immune from producing surprising and potentially misleading results even when permutations are used to construct a new model.  Very recent work has demonstrated that even though it seems quite unintuitive to think that a permuted feature could substantially improve predictive performance, such an outcome is possible in low signal settings \citep{Mentch2020}.  In light of this, we therefore see a \emph{condition}-and-relearn approach -- in which we both replace the feature values under investigation with samples generated conditional on the remaining features and construct a new model -- as essentially the gold standard for evaluating feature importance.  Put simply, we argue that procedures that involve conditioning on the remaining features (rather than simply permuting) or rebuilding the model are both substantial improvements over more classical ``permute-and-predict" approaches, but ideally procedures should do both.

This paper focusses specifically on permutation methods and variants of it, that can be applied to either the learned model, or the learning algorithm that generated it, in the spirit of staying within the algorithm class \citep{fisher2019all}.  However, these are not the only ways to try to understand the output of machine learning.  Our explanations of the poor behavior of permute-and-predict methods are traced to \citet{Hooker2007} where Brieman's variable importance methods were examined in the context of a functional ANOVA decomposition
\begin{equation} \label{eq:fanova}
f(\bm{x}) = f_0 + f_1(x_1) + \ldots + f_p(x_p) + \sum_{i < j} f_{ij}(x_i,x_j) + \ldots
\end{equation}
with the individual effects $f_{u}$, for a collection of covariate indices $u$ defined by averaging over the remaining features. Here, permutate-and-predict variable importance can be equated to Sobol indices \citep{sobol1993sensitivity} when all the features are independent \citep[see][for an early exploration of these tools for interpreting machine learning functions]{roosen1995visualization}. \citet{Hooker2007} proposed a generalization of this expansion to any distribution over the features, with the purpose of avoiding extrapolation. Similar results to the generalized functional ANOVA can be obtained by approximating a black box function in terms of additive models, described as {\em model distillation} in \citet{tan2018distill}. However, implementing these ideas is not as straightforward as the methods we discuss here.

Shapley values \citep{lundberg2017unified} have also become popular tools for assigning ``credit'' to covariates for any quantity, including predictive performance, or individual predictions. Some implementations of Shapley values exhibit the same permutation structure that we critique here and can be similarly misleading \citep{slack2020fooling}. Their implementation is also more complex than the methods we examine in this paper. Finally, we focus on procedures that are applicable to any machine learning method. Specific methods have their own diagnostics such as the split-improvement methods suggested in \citet{Friedman2001} that apply specifically to trees. We note that these methods can also be biassed towards features with more potential split points \citep{Strobl2007}, along with potential corrections in \citet{zhou2019,loecher2020unbiased,li2019debiased}.

In the following, we introduce a simple, and understandable model in Section \ref{sec:sim} used to illustrate the misleadingness of variable importance measures and diagnostic plots in Section \ref{sec:VI}. 
We provide an explanation of these results in terms of extrapolation behavior in Section \ref{sec:extrap} and suggest remedies in Sections \ref{sec:alt} and \ref{sec:GAMS}.

\section{A Simple Simulated Example} \label{sec:sim}

Here we set up a simple model that will allow us to be clear about the values that we ought to obtain from variable importance measures and which we can then contrast to those we actually find. Specifically, we propose a linear regression model based on 10 features $(x_1,\ldots,x_{10})$:
\begin{align}
y_i = x_{i1} & + x_{i2} +  x_{i3} + x_{i4} + x_{i5} + 0 x_{i6} \label{eq:sim}\\
& + 0.5 x_{i7} + 0.8 x_{i8} + 1.2 x_{i9} + 1.5 x_{i10} + \epsilon_i \nonumber
\end{align}
where $\epsilon_i \sim N(0,0.1^2)$ produces process noise. The variable importance methods we investigate here are not restricted to regression problems. However, in this setting we are able to relate all these diagnostic tools to the coefficients of the regression, providing a point of comparison. Our choice of noise variance is also deliberately small in order to make the behavior we observe clearer.

In addition to specifying the relationship between features and output, we need to provide a distribution for the $x_{ij}$. It is this that makes permute-and-predict methods misleading.  In our model, each of the features is marginally distributed as uniform on [0, 1]. They are generated independently with the exception of the first two $(x_1,x_2)$ which we make correlated via a Gaussian copula  \citep{Nelsen2007} with correlation parameter $\rho$. The strength of this correlation will play a large role in how strongly variable importance and other measures can mislead.

For a linear model such as \eqref{eq:sim} with covariates that have the same scale, variable importance can be obtained from the magnitude of covariates; see  \citet{Gregorutti2015} and Theorem \ref{thm:lin} below for a formal connection to the variable importance measures in \citet{randomforests}. Here we will be interested in the relative importance of $x_1$ and $x_2$ to the other features and we have chosen coefficients so that $(x_3,x_4,x_5)$ provide a reference for features with the same coefficient, $x_6$ has no influence and $x_1$ and $x_2$ have a clear importance ordering within $(x_7,\ldots,x_{10})$.

This example serves to isolate the reasons that the permute-and-predict methods can be misleading. We contend that permutation-based diagnostic tools are only misleading when one both employs flexible learning methods and has correlated features.  Neither, by themselves, are sufficient to induce large bias, although we note that the combination is in fact very common in practice. This necessity is illustrated in the following theorem that investigates the diagnostics returned by an estimated linear model.
\begin{theorem} \label{thm:lin}
For $f(x) = \hat{\beta}_0 + \sum_{j=1}^p \hat{\beta}_j x_j$ fit by least-squares
\begin{enumerate}
\item $E_{\pi} \mbox{VI}_j^{\pi} =  2 \hat{\beta}_j^2 \sum_{i=1}^N \left( x_{ij} - \bar{x}_j\right)^2$  where $E_{\pi}$ indicates the expectation over permutations and $\bar{x}_j$ is the average value of the $j$th feature.

\item $\mbox{PD}_j(x) = C_j + \hat{\beta}_j x$ where $C_j =  \sum_{j' \neq j} \beta_{j'} \bar{x}_{j'}$

\item $ICE_{i,j}(x) = C_{i,j} + \hat{\beta}_j x$ where $C_{i,j} = \sum_{j' \neq j} \beta_{j'} x_{ij'}$
\end{enumerate}
\end{theorem}
The first of these results is given in \cite{Gregorutti2015} and \cite{fisher2019all}; the latter two results follow in a direct fashion. A proofs is provided in the appendix.

Theorem \ref{thm:lin} is given in terms of an estimated linear model, but the analogous population result applies if these variable importance measures are applied to data generated from a linear model.  Theorem \ref{thm:lin} indicates that for linear models, permutation importance methods return the squared coefficient of the corresponding covariate, multiplied by the covariate's marginal sum of squares. When the covariates are standardized, this results in associating variable importance with the magnitude of the corresponding coefficient, approximately corresponding to common interpretations in linear models. We find that permutation importances do, in fact, recover importance ordering reliably when linear models are used as estimates in our simulation.  That breaks down, however, when random forests or neural networks are used to estimate the underlying function, where we see considerable bias in the ordering of importances relative to that given by the underlying (true) generating function.


In the simulations described below, for each choice of correlation parameter $\rho$ and sample size $N$, we learn both a random forest and a single layer neural network with 20 hidden nodes  (\texttt{randomForest} and \texttt{nnet} respectively in \texttt{R}). We then evaluated the
variable importance $\mbox{VI}^{\pi}$ as given above using the training data.
For random forests we also obtained the original out-of-bag importance measures $VI^o_j$ implemented in \texttt{importance}; these differ from $VI^{\pi}_j$ in being evaluated from out-of-bag residuals for each tree and are thus specific to bagging methods.


Additionally, we recorded the partial dependence of $f$ on $x_1$ and $x_2$ for each model as well as the ICE for 11 observations with the $i$th observation taken at $x_{i1} = x_{i2} = (i-1)/10$ and the remaining $x_{i3},\ldots,x_{i10}$ generated uniformly but kept constant across all simulations. These values were chosen so that the reference points were always in the bulk of the feature distribution.  We simulated each data set with associated learned function $f$, feature importances and plots, 50 times and report the average over these simulations in order to reduce variability in the results.


\section{Simulation Results} \label{sec:VI}

In order to ensure that scaling does not affect our results, we report the importance rank of each feature: for each simulation we order the importance of the features from least to greatest and record the position of the feature in this list.  That is, features deemed most important are given the highest rank.  This is commonly employed in screening methods where only the most important features are retained.

\begin{figure}
\centering
\includegraphics[height=7cm]{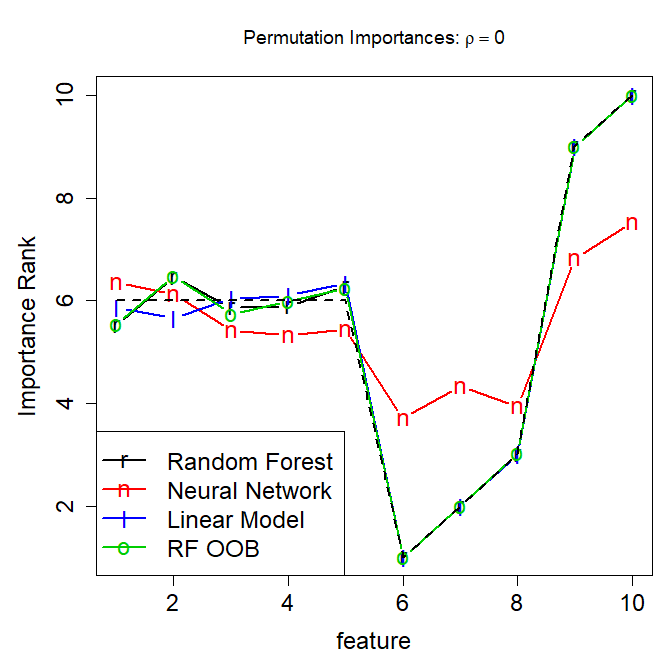}
\includegraphics[height=7cm]{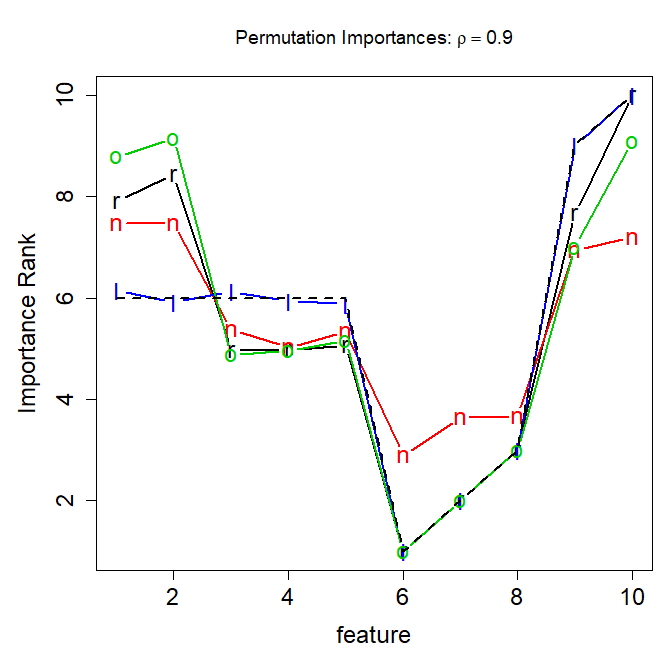}
\caption{Average variable importance rank (lowest to highest) computed using $VI^\pi$ on the training data for each feature over 10 models trained on simulated data of size 2000. Rank is given for random forests (r), neural networks (n) and linear models(l) as well as the out-of-bag variable importance for random forests (o).  Left: when all features are generated independently. Right: for $x_1$ and $x_2$ generated from a Gaussian copula model with correlation parameter $\rho = 0.9$. Dashed lines indicate the theoretical rank of the covariates. } \label{fig:PermImp}
\end{figure}

Figure \ref{fig:PermImp} shows the average importance rank of each of the 10 features obtained on the training set using the measures implemented in the \texttt{randomForest} package. We use 2,000 observations in each of 50 simulations and give results where $x_1$ and $x_2$ are related with a Gaussian copula with either correlation $\rho = 0$  or 0.9.  Note that in the independent case ($\rho=0$), random forests, neural networks and linear models all agree on the ordering of the covariates, including ties between $x_1$-$x_5$, which also corresponds to the results in Theorem \ref{thm:lin}.

However, when $\rho=0.9$, permutation-based methods applied to random forests and neural networks rank $x_1$ and $x_2$ as more important than $(x_3,x_4,x_5)$ and frequently more important than even $x_9$, which has a larger coefficient. However, in line with Theorem \ref{thm:lin}, linear models retained the same importance rank between the two correlation structures, an ordering which agrees with the known coefficients of the response function.

\begin{figure*}
\centering
\includegraphics[height=3.9cm]{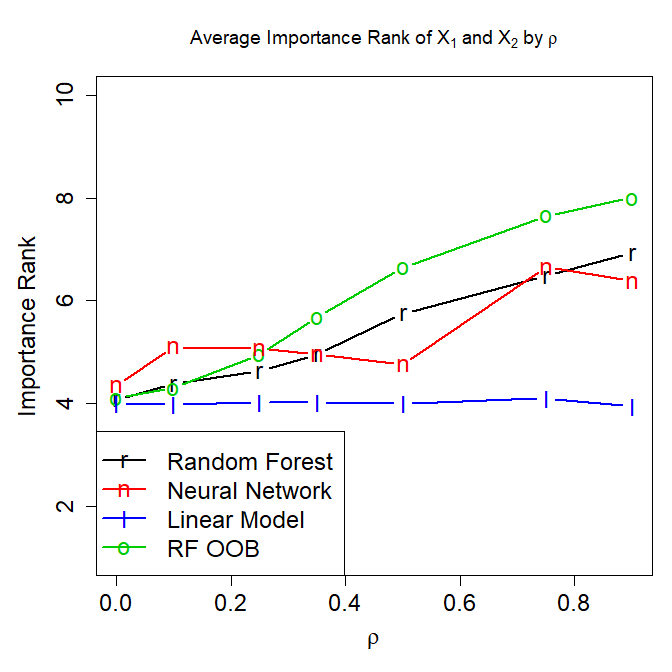}
\includegraphics[height=3.9cm]{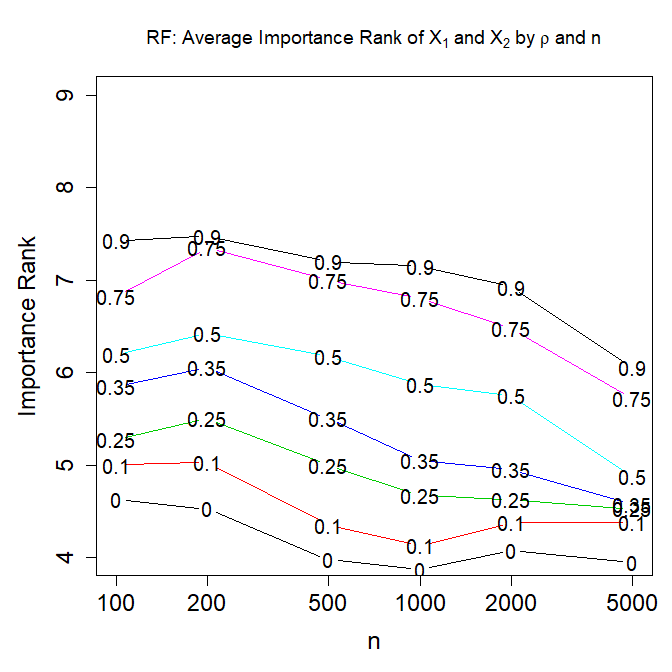}
\includegraphics[height=3.9cm]{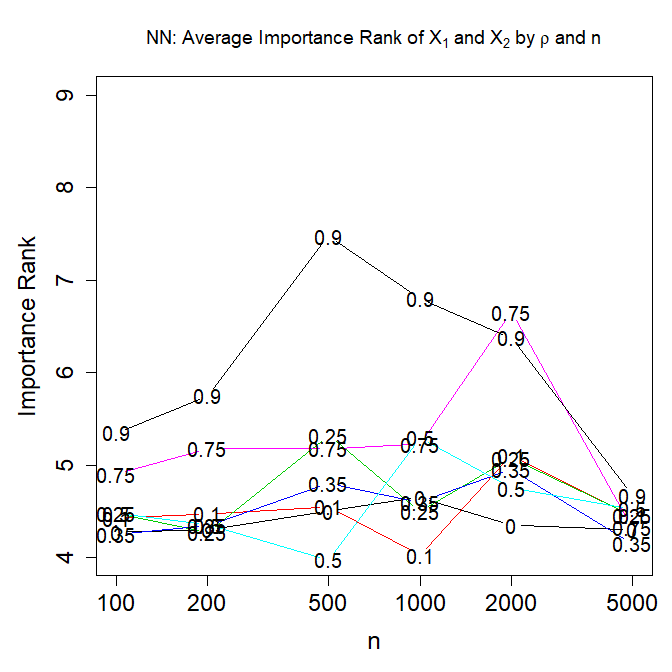}
\includegraphics[height=3.9cm]{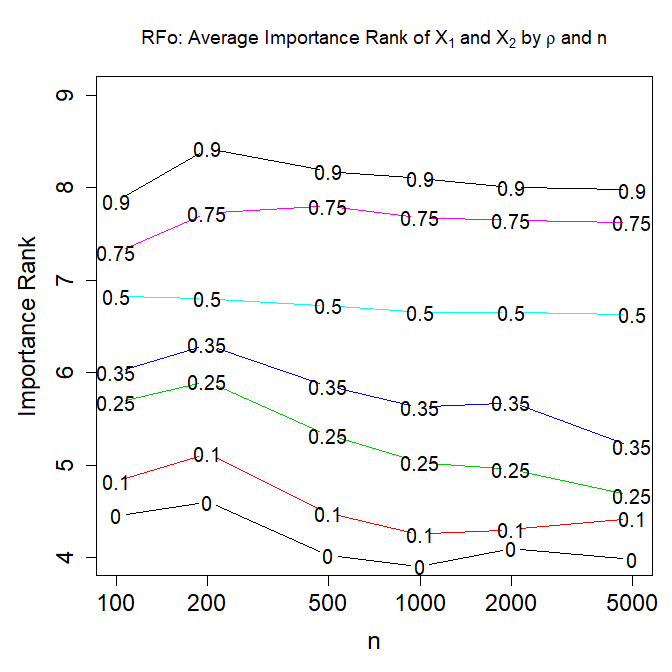}
\caption{Left: change in the average rank of $x_1$ and $x_2$ as correlation increases with $n=2000$. Remaining plots (left to right): average importance for each $\rho$ as a function of $N$ for fandom forests, neural networks, and random forests using OOB importance measures, respectively.  (True) theoretical rank should be 4. } \label{fig:C08}
\end{figure*}

In Section \ref{sec:extrap}, we explain this observation by noting that in permuting only 1 correlated feature, we break its relationship with the remaining correlated features resulting in evaluations of the function in regions which, by construction, have no data nearby; see Figure \ref{fig:design}. This may be viewed as being due to the very high correlation parameter employed, and the specific sample size.  In Figure \ref{fig:C08}, we examine the joint effect of sample size and correlation on the average rank given to $x_1$ and $x_2$. This time, we modify \eqref{eq:sim} so that each of $x_1$ and $x_2$ have coefficient 0.8, making them less relevant than $(x_3,x_4,x_5)$, and plot their average importance rank over $\rho \in  \{0,0.1,0.25,0.35,0.5,0.75,0.9\}$ and for each data set size $N \in (100,200,500,1000,2000,5000)$. These are averaged over 20 simulations to improve stability.

Here, we observe that for small $N$, nonzero correlation makes $x_1$ and $x_2$ appear to be more important than $(x_3,x_4,x_5)$, particularly for the out-of-bag importance measures, but that for small $\rho$ this effect is removed at larger $N$.  For large $\rho$, the importance rank decreases with $N$, though never below 5, the value indicated by Theorem \ref{thm:lin}.

\begin{figure*}
\centering
\includegraphics[height=3.9cm]{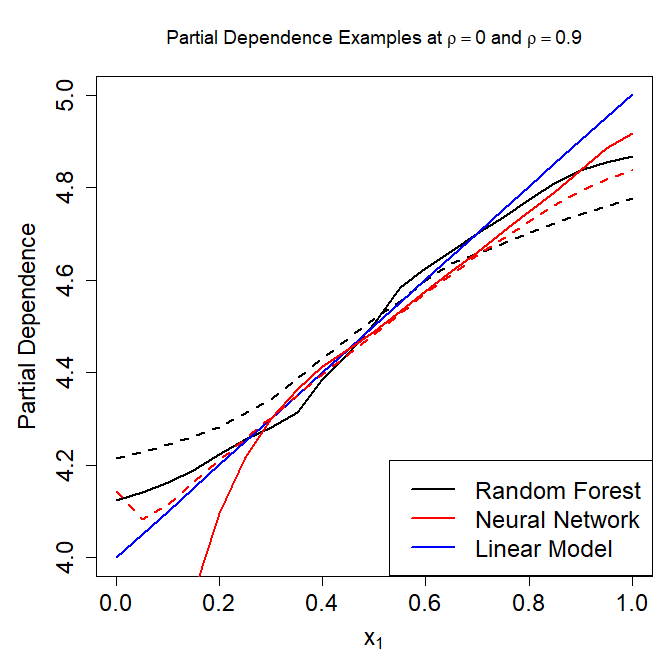}
\includegraphics[height=3.9cm]{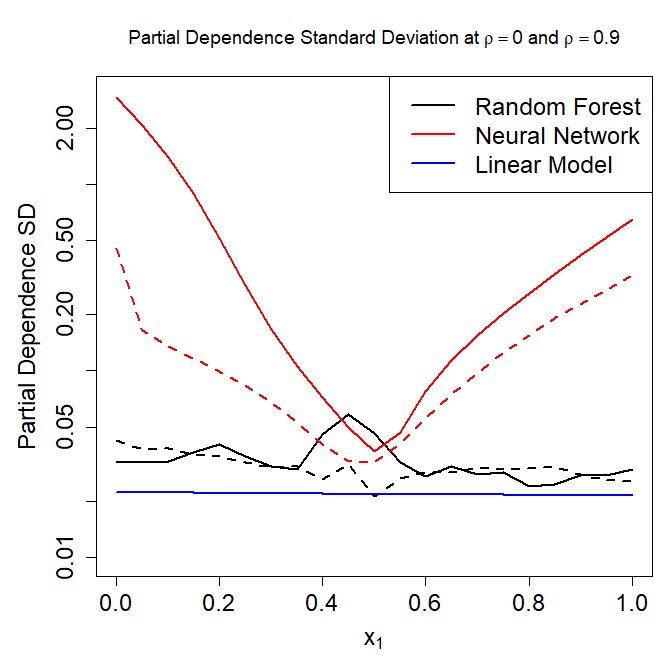}
\includegraphics[height=3.9cm]{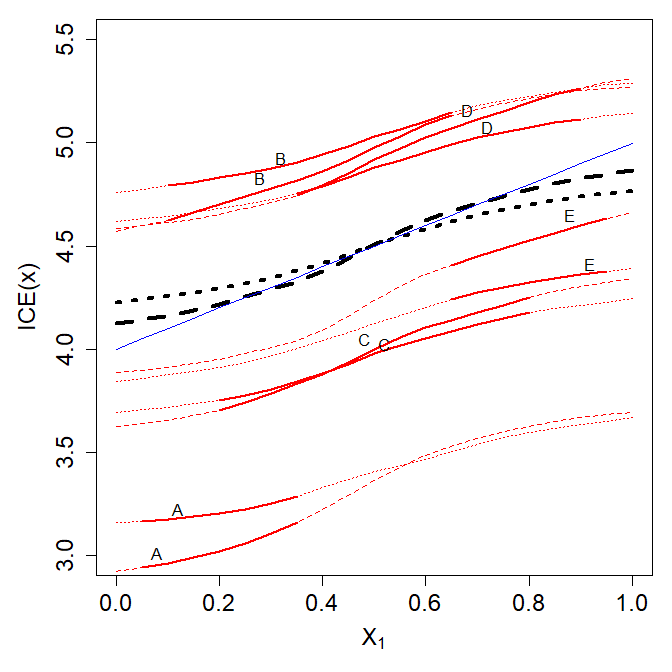}
\includegraphics[height=3.9cm]{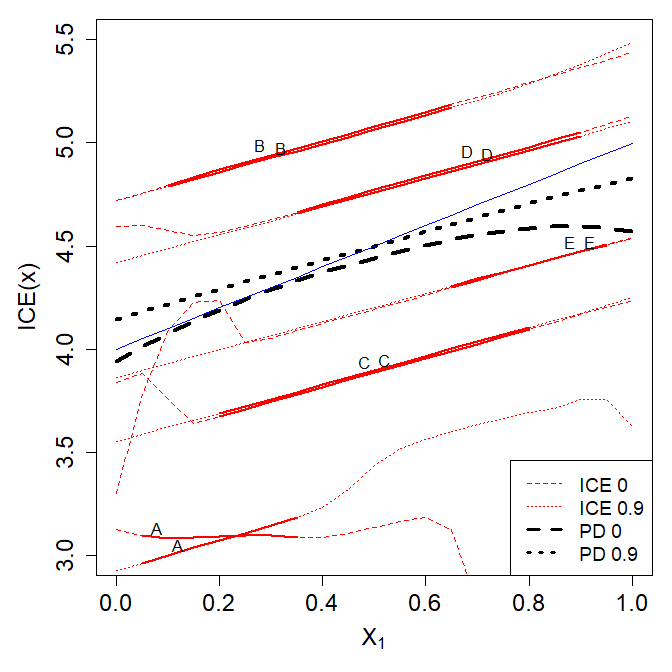}
\caption{Left plots: average partial dependence for 50 simulations both correlation parameters $\rho = 0$ (solid lines) and $\rho = 0.9$ (dashed) and standard deviation of the estimated partial dependence showing the increase in variability in neural networks as correlation increases, respectively.  Right plots: example ICE plots for $x_1$ for random forests (middle-right) and neural networks (far right). Dotted lines give predictions for simulations with $\rho = 0$, dashed for $\rho = 0.9$, letters indicate pairing. Solid portions of lines give the range of $x_1$ conditional on the remaining features.  Black lines indicate partial dependence functions; blue the underlying relationship. }   \label{fig:PD}
\end{figure*}

Figure \ref{fig:PD} gives the average partial dependence and ICE plots for $x_1$ for each of the models comparing those trained on independent feature distributions to those with $\rho=0.9$.  For random forests, we observe what appears to be less attenuation of the relationship due to edge effects when $x_1$ and $x_2$ are correlated, but note that these will still be compared to partial dependence functions for features where the edge effects are attenuated. Neural network partial dependence exhibits greater variability when the features are correlated. This is particularly evident in the ICE plots for neural networks where plots in the high correlation case diverge considerably from the underlying relationship.  Recall that each of these lines is an average of 50 replications, making the ICE for any individual curve highly variable.  The problematic behavior of these plots is more apparent in Figure \ref{fig:ICEExtrap} where a lower dimensional setting reduces both bias and variance.

\section{Extrapolation and Explanations} \label{sec:extrap}

The patterns observed in the results above have also been observed in, for example \cite{Tolosi2011}, although the periodic reappearance of these observations does not appear to have reduced the use of these types of diagnostics.  Furthermore, to this point, there has been relatively little explanation offered as to why this sort behavior occurs.

As noted in Theorem \ref{thm:lin}, for linear models with standardized features, permutation-based importances report the square of the feature coefficients. Moreover, where the features are not correlated, random forest permutation importances essentially provide the same ranking as permutation importances applied to linear models. The over-emphasis of correlated features is caused by random forests' (and neural networks') need to extrapolate in order to generate predictions at those permuted locations.

To illustrate this point, consider the model $y = x_1 + \epsilon$ so that $x_2$ does not influence the response. We generate $\epsilon \sim N(0,0.05)$ yielding a very high signal-to-noise ratio of 100/3 and learn a random forest based on 200 points with $(x_1,x_2)$ uniformly distributed but associated through a Gaussian copula with correlation 0.9.  This was repeated 100 times to stabilize the resulting estimates, and to allow us to investigate the between-simulation variability of our answers.

\begin{figure*}
\centering
\includegraphics[height=4.5cm]{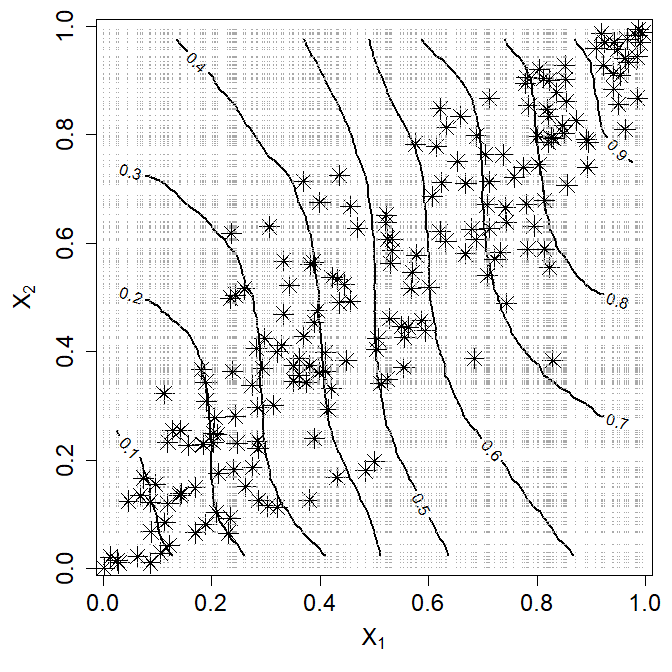}
\includegraphics[height=4.5cm]{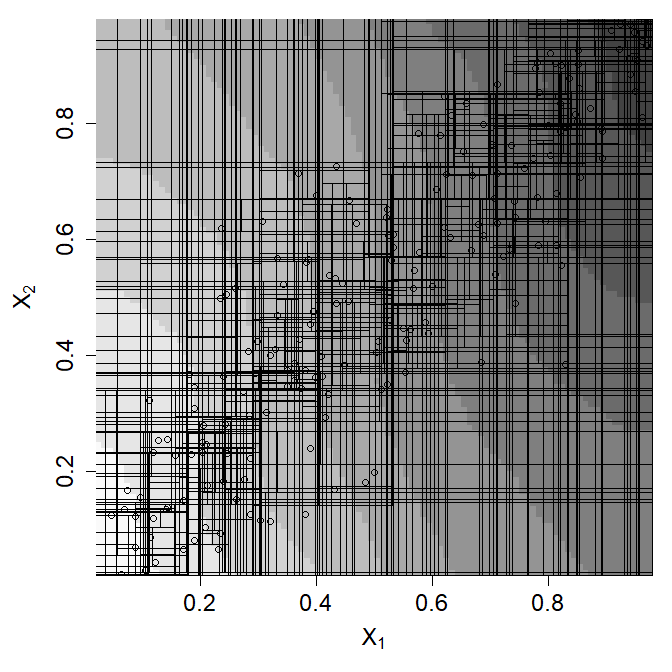}
\includegraphics[height=4.5cm]{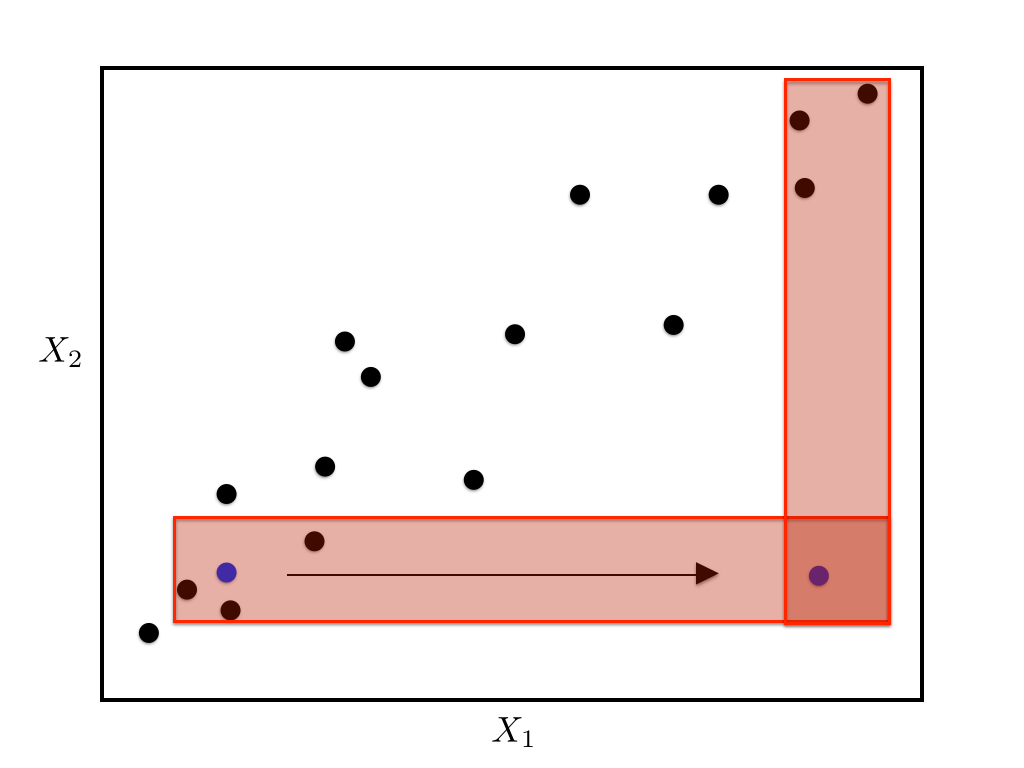}
\caption{Random Forest extrapolation behavior and variable importance. Left: contours from the average of 100 random forests trained on data given by '*' with response given by $x_1 + \epsilon$, dots give the locations at which the forests are queried when calculating permutation-based variable importance. Middle: The boundaries between leaves for 10 example trees in the forest: when the forest is extrapolating the trees are local averages  based either on $x_1$ or $x_2$.  Right: an illustration of potential nearest neighbours of a query point used to determine variable importance.
} \label{fig:design}
\end{figure*}

The left panel Figure \ref{fig:design} plots the generated $(x_1,x_2)$ points, the points used to assess permutation importance for $x_2$ and the contours of the learned random forest.  Here we observe two things:
\begin{enumerate}
\item Although the true contours are given by vertical lines, the random forest only approximates these contours within the convex hull of the data. Outside this, the contours become more diagonal.

\item While the data distribution is concentrated along the diagonal, the points used to evaluate permutation importance fill out the unit square.
\end{enumerate}
Thus, much of the evaluation of permutation importance is based on values of $\bm{x}$ that are far from the training data, where random forests are (unsurprisingly) poor at mimicking the underlying data generating function.

The middle plot of Figure \ref{fig:design} overlays a shaded contour with the splits obtained from 10 example trees selected at random from our forests.  Here we observe that at the top left and bottom right corners, individual tree predictions are obtained either by splits that proceed horizontally from or vertically from the data distribution at the adjacent corners.  Thus, as discussed, each tree will predict a local average from one of the two corners.

The right hand plot of Figure \ref{fig:design} makes this reasoning more explicit, in which the permuted value $x_{i1}^{\pi}$ moves the query from the original data in the bottom left, to $\bm{x}^{\pi}$ in the bottom right of the plot where there are very few observations. In tree-based methods, predictions from each tree at $x$ take the form
  \begin{equation}
    \hat{f}(x) = \sum_{i=1}^{N} w_i y_i \quad \text{ where } w_i =
    \begin{cases}
      1/|L(x)|, & \text{if } x \in L(x) \\
      0 & \text{otherwise}
    \end{cases}
  \end{equation}
where $L(x)$ denotes the set of observations falling in the same leaf as $x$.  The collection of response values $y_i$ for which the corresponding weight $w_i$ can be non-zero is a subset of all training observations and such observations are referred to as the \emph{potential nearest neighbors} (pNN) of $x$.

For correlated features, the potential nearest neighbours of $\bm{x}^{\pi}$ include those whose values are far from the original $\bm{x}$ in both coordinates; indeed this happens with substantial frequency among the trees in our forest.  In these trees, the permuted prediction $\hat{f}(x^{\pi})$ is likely to be far from the original prediction $\hat{f}(x)$, causing a large perceived importance of the feature $x_1$ even when it is irrelevant.   By contrast, when the observations are more uniformly distributed over the feature space, the pNN's of $\bm{x}^{\pi}$ will be localized around it, and will have very similar values of $x_2$ to the original point $\bm{x}$ making $f(\bm{x}^{\pi})$ a reasonable comparison to $f(\bm{x})$.

In fact, almost any of the data points in the right-most plot of Figure \ref{fig:design} are pNNs of $\bm{x}^{\pi}$. This can be seen by examining the set of rectangles that reach both the data distribution and the bottom-right corner of the plot. Moving along one edge or other, however, the geometry of forming a rectangle that encompasses both a small number of data points and the prediction restricts potential nearest neighbours to be those with larger values of both $x_1$ and $x_2$ (or smaller values for points above the diagonal).


In Figure \ref{fig:design}, this argument explains the shift in contours away from the observed data and an increase in the importance measure for $x_2$. However it is not sufficient to explain the joint increase in importance for {\em both} $x_1$ and $x_2$ when they have symmetric effects in the simulations in Section \ref{sec:sim}. To explain this effect, we also observe that the concentration of observations along the diagonal increases the over-all signal jointly attributed to them, increasing the number of splits given to at least one of these two covariates. 

\begin{figure*}
\centering
\includegraphics[height=3.9cm]{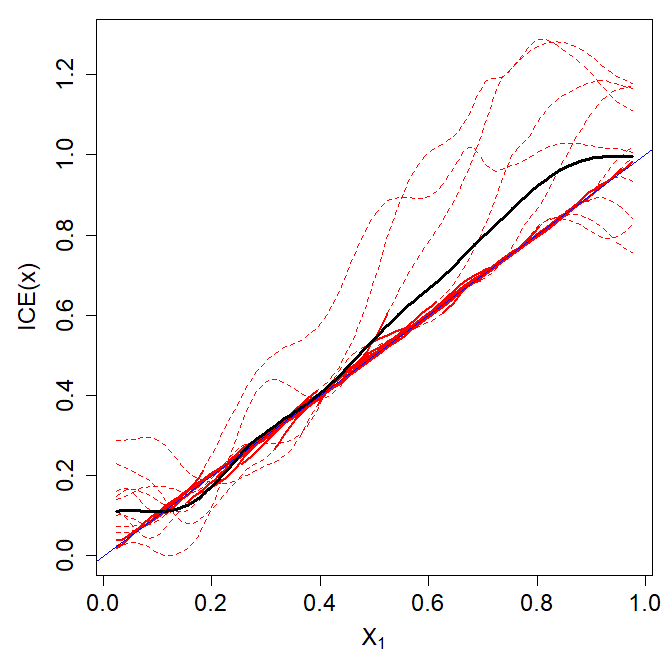}
\includegraphics[height=3.9cm]{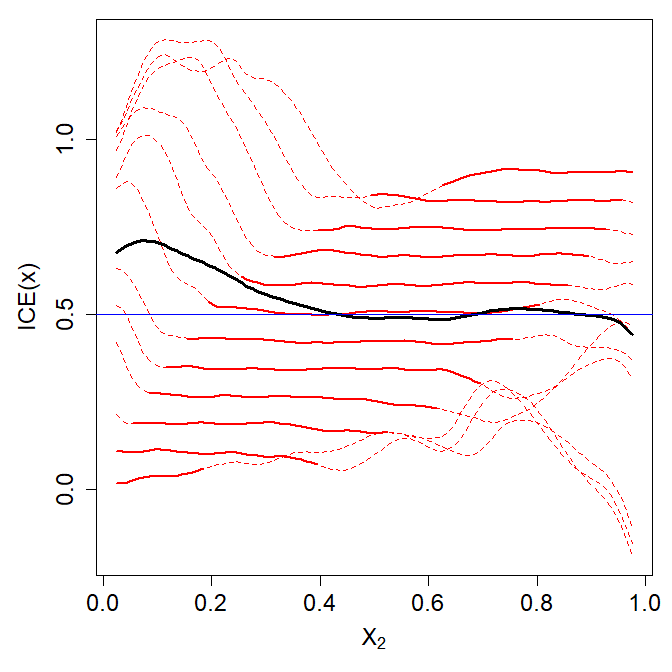}
\includegraphics[height=3.9cm]{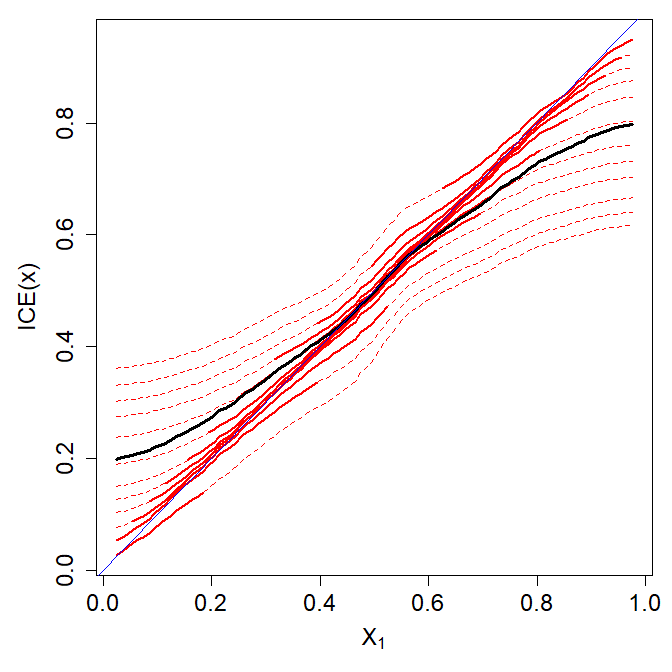}
\includegraphics[height=3.9cm]{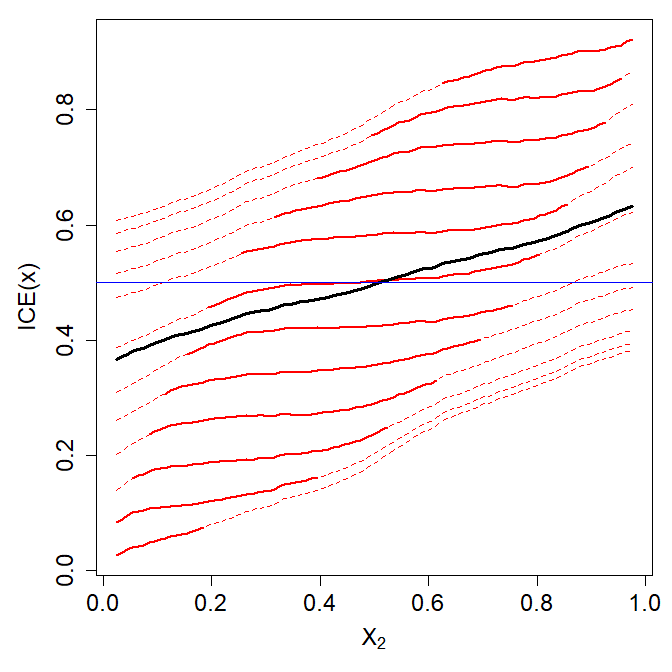}
\caption{Example ICE plots for models trained on data generated from $y=x_1 + \epsilon$. Left pair: neural network. Right pair: random forest. Thick black lines give partial dependence, thin blue lines indicate the theoretical model.} \label{fig:ICEExtrap}
\end{figure*}

Figure \ref{fig:ICEExtrap} plots the average PD and ICE functions for each of random forests and neural networks in the same manner as Figure \ref{fig:PD}. Here, the effect of extrapolation is made even more evident.  ICE plots reflect the underlying relationship reasonably well in regions that are supported by the data, but can be very different outside of that. In the case of neural networks, these exhibit greater variability; in random forests, extrapolation-induced inflation of importance measures is a larger source of concern, suggesting more significance for $x_2$ than we would expect.

\begin{figure}
\centering
\includegraphics[height=7cm]{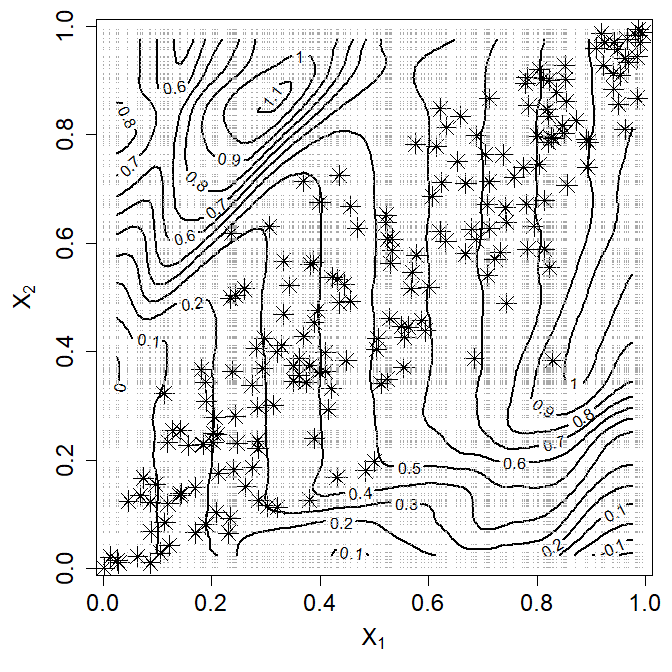}
\includegraphics[height=7cm]{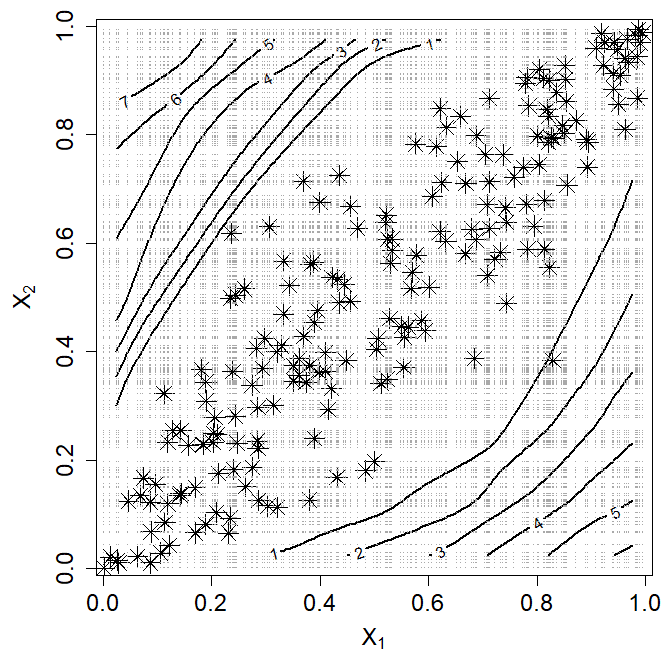}
\caption{Left: contour plot of the average of 100 neural networks trained on correlated data with response $x_1 + \epsilon$. Example data is given by `*' with dots indicating the points used to evaluate variable importance. Right: the standard deviation between the 100 neural networks. } \label{fig:nn.extrap}
\end{figure}

In contrast to random forests, our characterization of neural network extrapolation is simply in terms of increased variability. When fitting high-dimensional polynomials, the phenomenon of observing large oscillations outside the range of the data is termed {\em Gibbs effects} and we suspect that a similar phenomenon occurs here.  Figure \ref{fig:nn.extrap} presents a contour plot of the average of 100 20-hidden-node neural networks trained on the same data as above along with the standard deviation between these models. Here, the lack of reliability in permutation importance and in partial dependence plots can be explained by the large increase in variance as we move away from the data distribution, allowing fluctuations in predictions to increase apparent variable importance.

\section{Variable Importance Alternatives} \label{sec:alt}
As has been noted, these effects of correlation on variable importance have been observed in various forms by several authors in recent years.  Fortunately, there have also been several alternatives proposed, generally in line with one of two ideas:
\begin{enumerate}
\item Permuting (or otherwise generating) new values of feature $j$ based on its distribution conditional on the remaining features.  This general idea seems to have first been suggested in \citet{Strobl2008} but similar schemes have since appeared in numerous other works including \citet{Tuv2009}, where the authors utilize tree ensembles to generate predictively non-redundant feature sets, and \citet{Wu2007} where the original dataset is augmented with \emph{pseudovariables} in order to help tune the size and complexity of linear models.  More recently, \citet{Barber2015} and \citet{Candes2018} developed the ``knockoff" framework whereby original features are either swapped out for randomized replacements to test their importance via conditional randomization tests or the feature space is augmented with null copies of the original features after which a knockoff filter is applied to eliminate non-important features.

    \citet{fisher2019all} similarly examines model class reliance using conditional permutations. The notion of model class reliance examines how widely an importance measure can vary while maintaining close-to-optimal predictive accuracy. Although we do not examine these methods here, the results we present suggest that this conditional version should be preferred.

\item Removing feature $j$ from the data and examining the drop in accuracy when a new model is learned without it. This leave-one-covariate-out (LOCO) framework is well studied within classical statistical methods, (e.g.\ \citet{Lehmann2006}) and suggested more generally in \citet{Lei2018} in the context of conformal inference.
\end{enumerate}
These methods can also be combined. \citet{Mentch2016} examined the change in prediction between two random forests, one trained on the original data, the other trained on data in which feature $j$ was permuted.  Here the permutation served simply to make the feature irrelevant to the prediction and the resulting models were both evaluated within the bulk of the training distribution. As the authors note, the choice to permute, rather than drop, a feature was made in order to maintain the same complexity in the fitting procedure and thus make the two predictions statistically comparable. 

In fact, these methods all produce similar variable importance scores. In our simulation in Section \ref{sec:sim}, we also computed the following variable importance measures that combine these ideas:
\begin{description}
\item[Conditional Variable Importance] measured by creating $X^{c,j}$ in which $x_{ij}$ is simulated conditional on the remaining features $x^{c,j}_{ij} \sim x_{ij} | \bm{x}_{i,-j}$ and measuring
    \[
    \mbox{VI}^C_j = \sum_{i=1}^N L(y_i, f(\bm{x}^{c,j}_i)) - L(y_i,f(\bm{x}_i))
    \]
    We note that for the Gaussian copula model used here, the distribution of $(x_{ij}|\bm{x}_{i,-j})$ can be explicitly computed. Note that the idea of conditional permutation importance for random forests proposed in \cite{Strobl2008} falls into this category, as does the notion of conditional model reliance in \cite{fisher2019all} as described above. In \cite{Strobl2008}, the authors generate $X^{c,j}$ for each tree by permuting $X_j$ within cells of a partition generated based on all cutting points related to $X_{-j}$ in that tree.

\item[Dropped Variable Importance] obtained by the increase in training error when learning a model $f^{-j}(\bm{x})$ from the data $\bm{y},X_{-j}$ that does not include the $j$th feature
    \[
    \mbox{VI}^D_j = \sum_{i=1}^N L(y_i, f(\bm{x})_i) - L(y_i,f^{-j}(\bm{x}_i)).
    \]
    This is equivalent to the LOCO methods explored in \cite{Lei2018}.

\item[Permute-and-Relearn Importance] obtained by permuting the features and learning $f^{\pi,j}$ from $(\bm{y},X^{\pi,j})$ giving
    \[
    \mbox{VI}^{\pi L}_j = \sum_{i=1}^N L(y_i, f(\bm{x}_i)) - L(y_i,f^{\pi,j}(\bm{x}_i)).
    \]
    This was the approach taken in \cite{Mentch2016} in which distributional results for random forests were used to assess the statistical significance of $f(\bm{x}) - f^{\pi,j}(\bm{x})$.

\item[Condition-and-Relearn Importance] in which a new model $f^{c,j}(\bm{x})$ is learned from the data $(\bm{y},X^{c,j})$ and we measure
    \[
    \mbox{VI}^{CL}_j = \sum_{i=1}^N L(y_i, f(\bm{x}_i)) - L(y_i,f^{c,j}(\bm{x}_i))
    \]
\end{description}
 These measures all necessarily entail additional computational costs: requiring training a new model, and/or estimating, and simulating from, the distribution of $(x_{ij}|\bm{x}_{i,-j})$. In our simulation this distribution can be computed analytically, but that will rarely if ever be the case in practice.

 Using these measures changes the estimand of variable importance. In the case of least-squares loss these do all target the same quantity:
 \begin{theorem} \label{thm:oracle}
    Let $\bm{x}_j^*$ be obtained by replacing $x_j$ with $x_j^*$ where $x_j^* \perp y |\bm{x}_{-j}$ and $f(\bm{x}) = E(y|\bm{x})$.  Then the minimum least squares predictor of $y$ from $\bm{x}^*$ is
    \[
    f_{-j}(x^*) = \int f(\bm{x}) p(x_j|\bm{x}_{-j}) dx_j
    \]
    and the corresponding variable importances for this $f$ are given by
    \[
    \mbox{VI}^C_j = \mbox{VI}^D_j = \mbox{VI}^{\pi L}_j = \mbox{VI}^{CL}_j = E_{y,\bm{x}} ( f(\bm{x}) - f_{-j}(\bm{x}_{-j}) )^2.
    \]
 \end{theorem}
 Note that outside of squared error, the predictor minimizing loss when predicting from $\bm{x}^*$ may no longer be $f_{-j}(\bm{x}_{-j})$ and thus the variable importance definitions may differ.

\begin{figure}
\centering
\includegraphics[height=4.5cm]{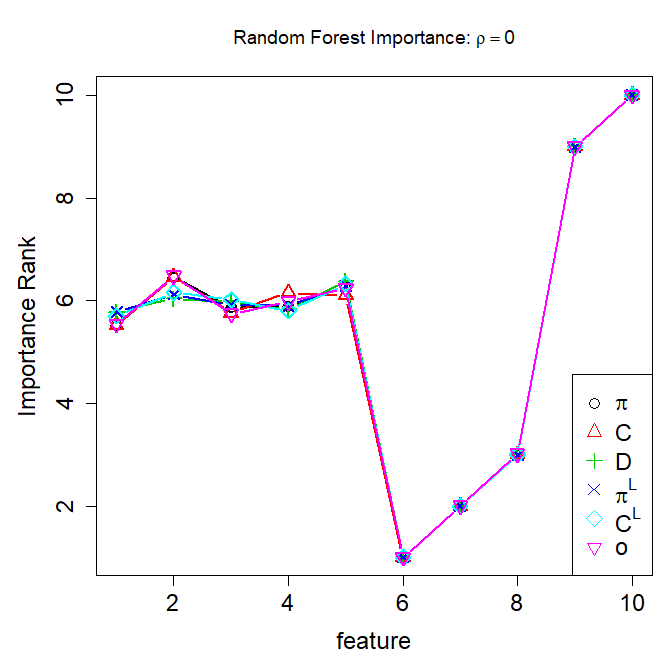}
\includegraphics[height=4.5cm]{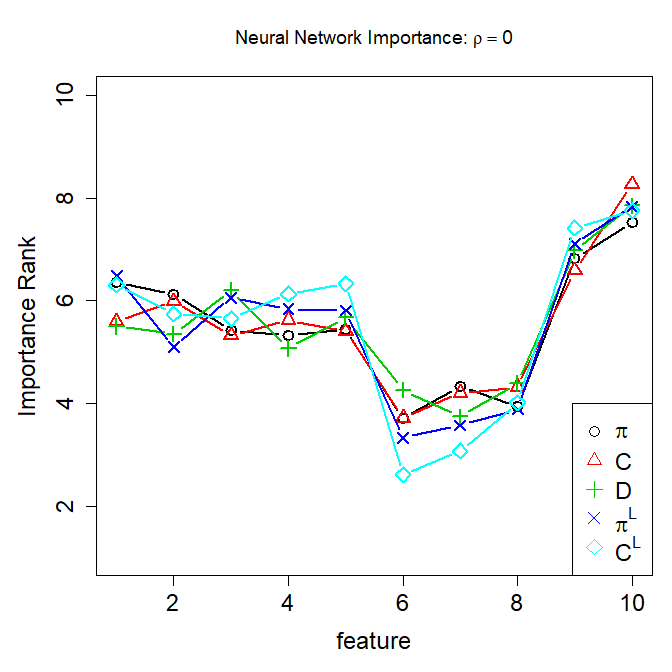}
\includegraphics[height=4.5cm]{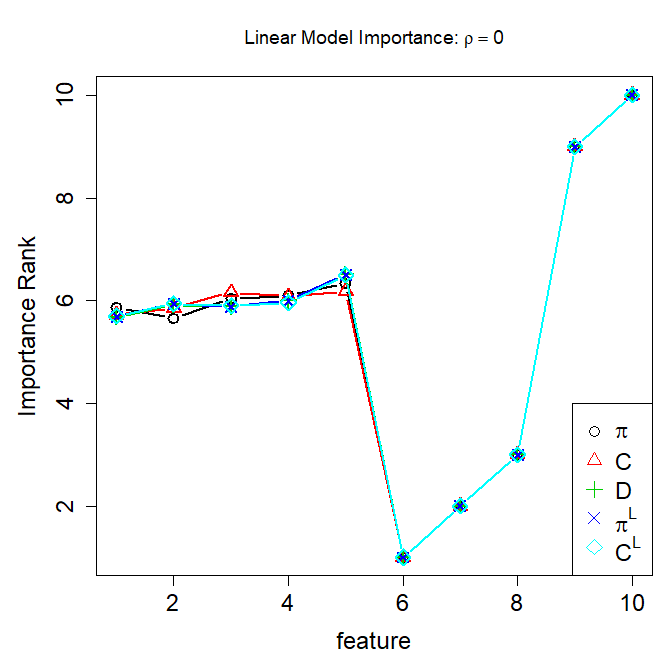} \\
\includegraphics[height=4.5cm]{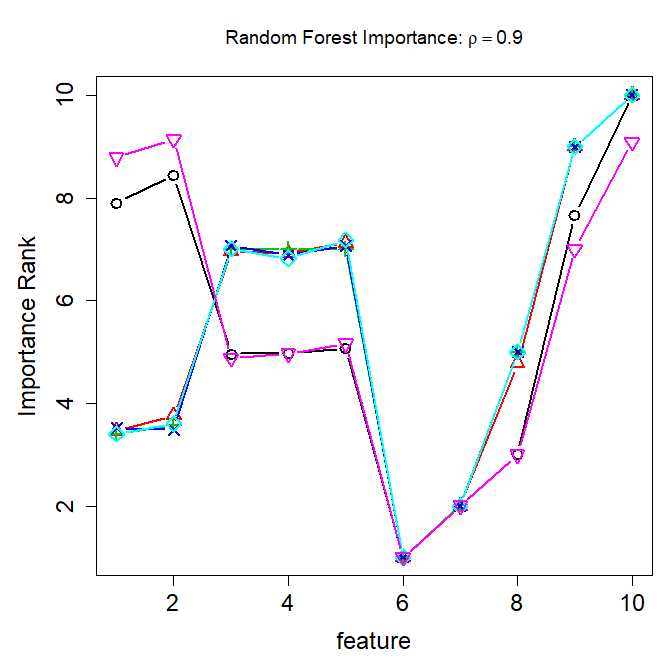}
\includegraphics[height=4.5cm]{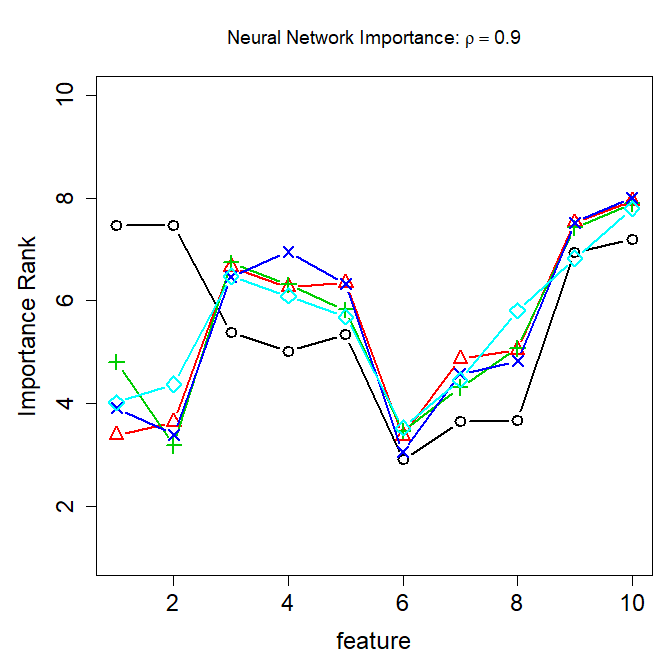}
\includegraphics[height=4.5cm]{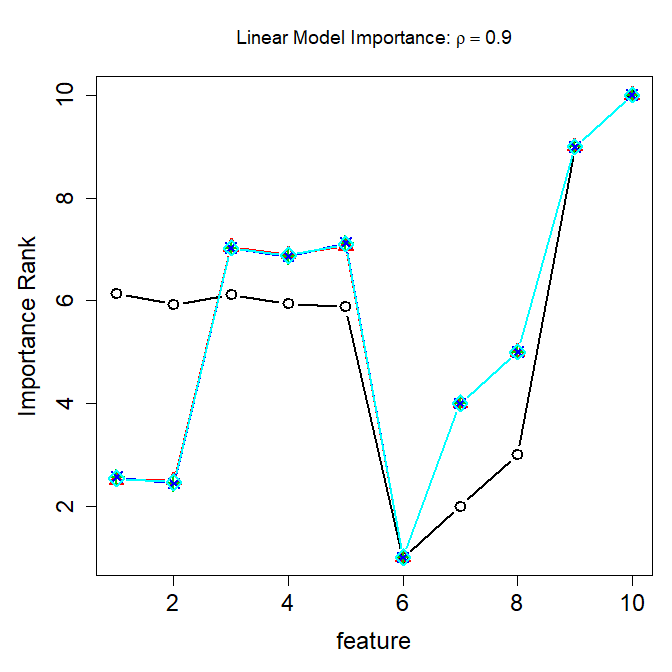}
\caption{Average rank of feature importance computed by various measures for random forests (left), neural networks (middle) and linear models (right) based on 10 simulated data sets of size 200. Features were generated independently (top) and with $x_1$ and $x_2$ generated from a Gaussian copular with correlation parameter $\rho = 0.9$ (bottom). } \label{fig:NoPermImp}
\end{figure}

As we can see in Figure \ref{fig:NoPermImp}, these measures also all agree (to within Monte Carlo variability) on the ordering of covariate importance across all models. Unlike the permute-and-repredict measures, they reduce the importance of $x_1$ and $x_2$ when these become more correlated. For LOCO methods, this can be explained intuitively by $x_1$ being able to account for the some of the signal coming from $x_2$. For conditional importance measures, it is associated with the distribution $(x_2|x_1)$ having a much smaller variance than the marginal distribution of $x_2$ (see, for example, the points in Figure \ref{fig:nn.extrap}).

This is, in fact, what we should expect. For estimated linear models, Theorem \ref{thm:drop.equal} states that these all have approximately the same expectation:
\begin{theorem} \label{thm:drop.equal}
Let $f(x) = \hat{\beta}_0 + \sum_{j=1}^p \hat{\beta}_j x_j$ be a model fit via least-squares with linear dependence between features so that $E(x_{ij}|\bm{x}_{-j}) = \gamma_{0j} + \bm{x}_{i,-j} \bm{\gamma}_j$. Let $\delta_{ij} = x_{ij} - \hat{\gamma}_0 -  \bm{x}_{-j}\bm{\gamma}_j$ be the least-squares residuals to predict $x_{ij}$.  Then
\begin{align*}
&  2 \mbox{VI}^D_j =  E_{\pi} \mbox{VI}^{\pi L}_j  = E_{x_{ij}^{c,j}} \mbox{VI}^{CL}_j  = 2  \beta_j^2 \sum_{i=1}^{N} \delta_{ij}^2 + o(1) \\
& E_{x_{ij}^{c,j}} \mbox{VI}^C_j  =   2 \beta_j^2  \sum_{i=1}^{N} \mbox{var}(x_{ij}|\bm{x}_{-j}).
\end{align*}
\end{theorem}
Here the addition of $o(1)$ comes from estimating coefficients with a random covariate matrix when permuting a feature; when the importances are normalized by $N$ this becomes $o(1/N)$ and is generally small.

All of these methods exhibit the compensation effect; reducing the importance of correlated features relative to their coefficients. Theorem \ref{thm:drop.equal} also suggests a way to correct for this effect by dividing by the appropriate quantity, if desired.   However, these results are only exact for linear models and the most reliable diagnostic that we know of for this is to examine the effect of jointly removing or permuting pairs of features, and then re-learning. An extension of Theorem \ref{thm:lin} yields a joint importance of $\sum_{i=1}^N ( \beta_1 (x_{i1}-\bar{x}_1) + \beta_1 (x_{i2}-\bar{x}_2))^2$, which is not recoverable from any of univariate importance measures discussed in this paper.


%
%
%

A further set of approaches to variable importance use the behavior of $f(\bm{x})$ close to values of $\bm{x}$ in the data.  Saliency \citep{Simonyan2013} obtains derivatives with respect to $\bm{x}$, while LIME \citep{Ribeiro2016} approximates this with a LASSO method localized to the point of interest by a kernel; Shapley values can also be used for individual predictions \citep{lundberg2017unified}. By querying the value of $f(\bm{x})$ close to the features of interest, these should not suffer from extrapolation; although see  \citet{slack2020fooling} for an example of distorting both SHAP and LIME by modifying $f(\bm{x})$ when $\bm{x}$ is not in the feature distribution.  However, local explanations  do not generally provide the same global-level picture of behavior that permutation methods purport to. For example, if $f(x)$ essentially encodes a threshold with a transition point that is not close to observed data, we will see low saliency at all observed points, even though there may be a large range of predictions. Thus, while these kinds of localized explanations may help provide insights into the features affecting particular predictions, they lack the ability to offer an overall summary of a feature's contribution to the model.

\subsection{Ordering vs.\ Testing:  A Further Word of Caution}
As discussed throughout this paper, while the permute-and-repredict measures of variable importance can be highly misleading in the presence of correlated features, several alternative measures such as the four mentioned above can alleviate these biases by refitting the models under investigation, leading to importance orderings that are more intuitive, stable, and consistent.  We caution here, however, that while such measures can correct improper variable importance orderings, 
they may not be sufficient to employ variable importances within hypothesis tests.

Formal hypothesis tests for variable importance in random forests were first proposed in \cite{Mentch2016} and an alternative nonparametric test with potential to operate in big-data settings was recently proposed in \cite{Coleman2019}. A similar model-agnostic variant was recently proposed in \cite{williamson2020} Rather than assign a particular measure of variable importance, these tests assume that such a measure already exists and evaluate whether a given feature or collection of features is significantly more important than could be expected by random chance.  More formally, these tests partition the features into two sets $\bm{x}_1$ and $\bm{x}_2$ and consider a null hypothesis that can be written generically as
\[
H_0:  \text{Error} \left[ RF(\bm{x}_1,\bm{x}_2) \right] = \text{Error} \left[ RF(\bm{x}_1,\bm{x}_{2}^{*}) \right]
\]
and evaluated on an external test set.  Here, $RF(\bm{x}_1,\bm{x}_2)$ denotes a random forest constructed on the original features $(\bm{x}_1,\bm{x}_2)$ and $RF(\bm{x}_1,\bm{x}_{2}^{*})$ denotes a random forest constructed on the same original features $\bm{x}_1$, but where $\bm{x}_2$ is either dropped or replaced by a non-important, randomized substitute such as a permutation or a new sample conditional on $\bm{x}_1$.  Note that these correspond exactly to the importance measures described in the preceding subsection.

The intuition behind these tests is that if the features in $\bm{x}_2$ are important, then they should make a meaningful contribution to the predictive accuracy of the model beyond what can be obtained by using $\bm{x}_1$ alone.  In that case, we should expect that $\text{Error} \left[ RF(\bm{x}_1,\bm{x}_{2}^{*}) \right] > \text{Error} \left[ RF(\bm{x}_1,\bm{x}_2) \right]$ and we can thus reject $H_0$ and conclude that $\bm{x}_2$ is of \emph{significant} importance.

In recent work, however, \cite{Mentch2020} showed that the manner in which $\bm{x}_{2}^{*}$ is altered can have a substantial impact on the performance of the aforementioned hypothesis tests.  In particular, the authors demonstrate that there are settings in which the ``Permute-and-Relearn" approach can identify $\bm{x}_2$ as predictively significant, even when those features are completely independent of the response, conditional on $\bm{x}_1$.

As a demonstration of this, consider again the same linear regression model
\begin{align}
y_i = x_{i1} & + x_{i2} +  x_{i3} + x_{i4} + x_{i5} + 0 x_{i6} \nonumber \\
& + 0.5 x_{i7} + 0.8 x_{i8} + 1.2 x_{i9} + 1.5 x_{i10} + \epsilon_i \nonumber
\end{align}
utilized in previous sections, but where observations are sampled independently from $N(0,\Sigma)$ where the $(i,j)^{th}$ entry of $\Sigma$ is of the form $\rho^{|i-j|}$ so as to induce correlation among these features.  The additional error $\epsilon_i \sim N(0,\sigma^2)$ and $\sigma$ is chosen to produce a desired signal-to-noise ratio (SNR) in the dataset.  Now suppose that in addition to these 10 original features, we also consider including $q$ additional noise features.  Each of these noise features is sampled from a standard normal and then correlated with one of the original 10 features -- selected uniformly at random -- at a strength of 0.7 so that conditional on those original 10 features, the $q$ additional features are independent of the response $y$.  We then carry out the testing procedure prescribed in \cite{williamson2020} on both bagging and random forests, treating $\bm{x}_2$ as the set containing the $q$ additional noise features at varying SNRs and sizes of $q$.

\begin{figure}
\centering
\includegraphics[width = 0.31\textwidth]{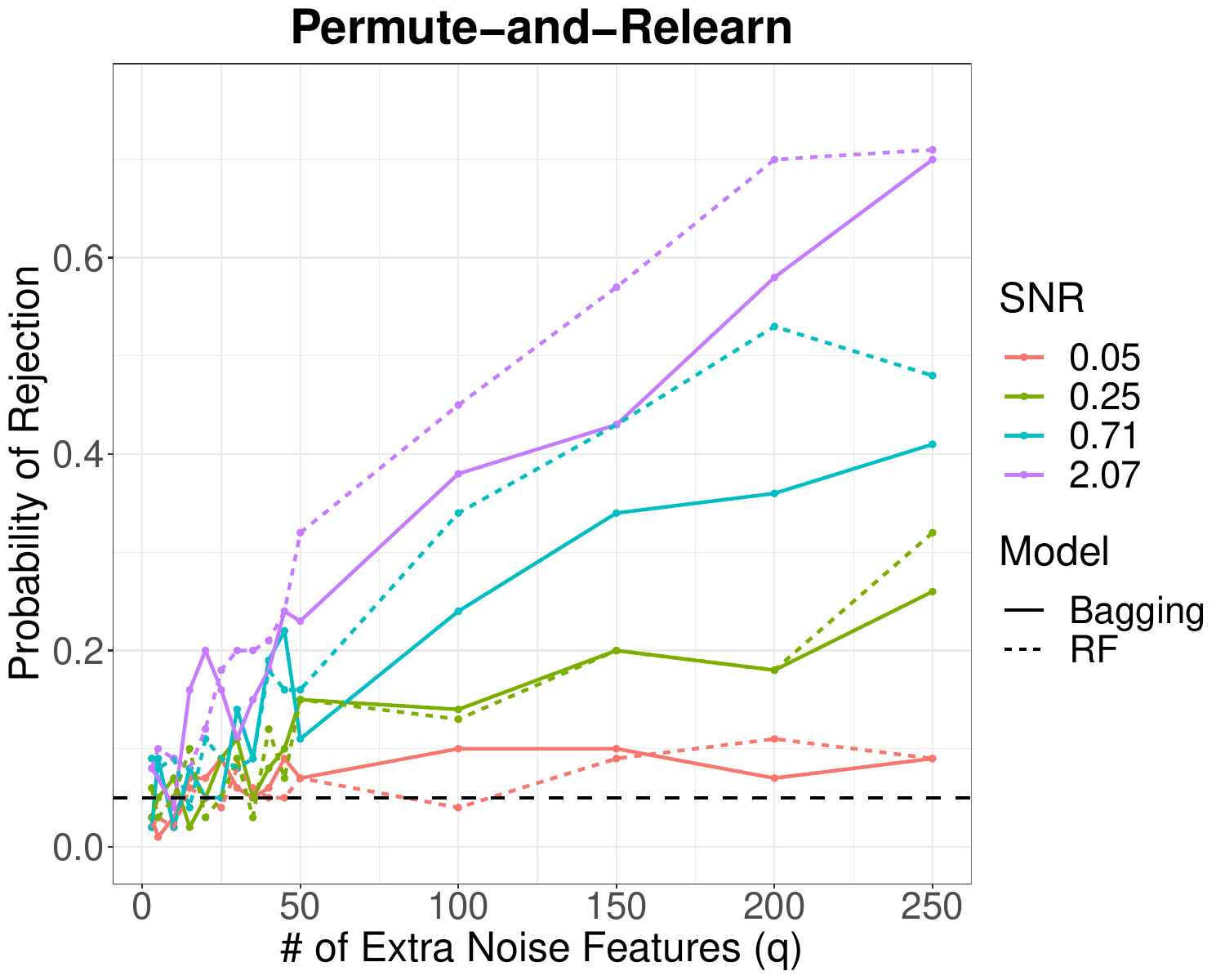}
\includegraphics[width = 0.31\textwidth]{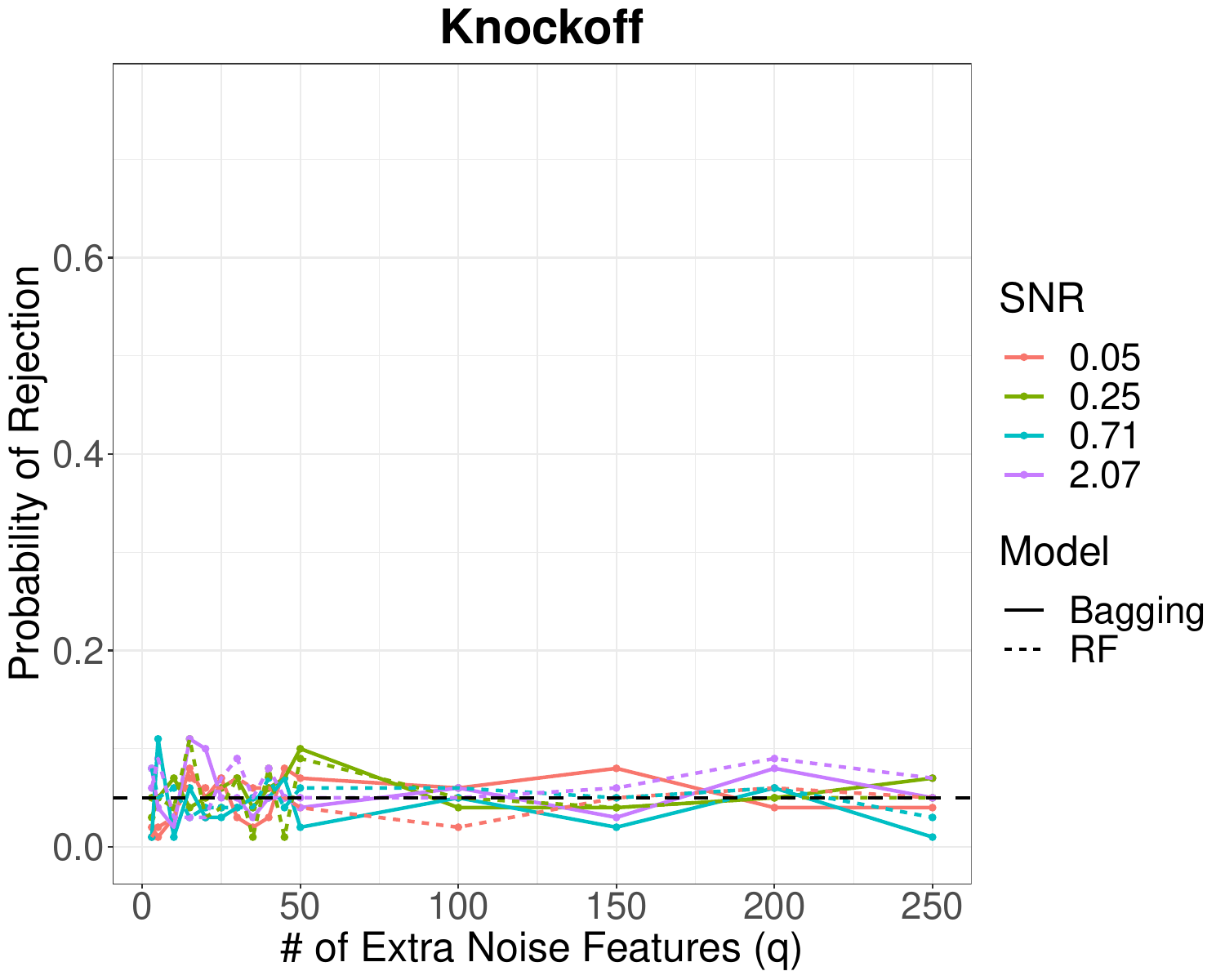}
\includegraphics[width = 0.31\textwidth]{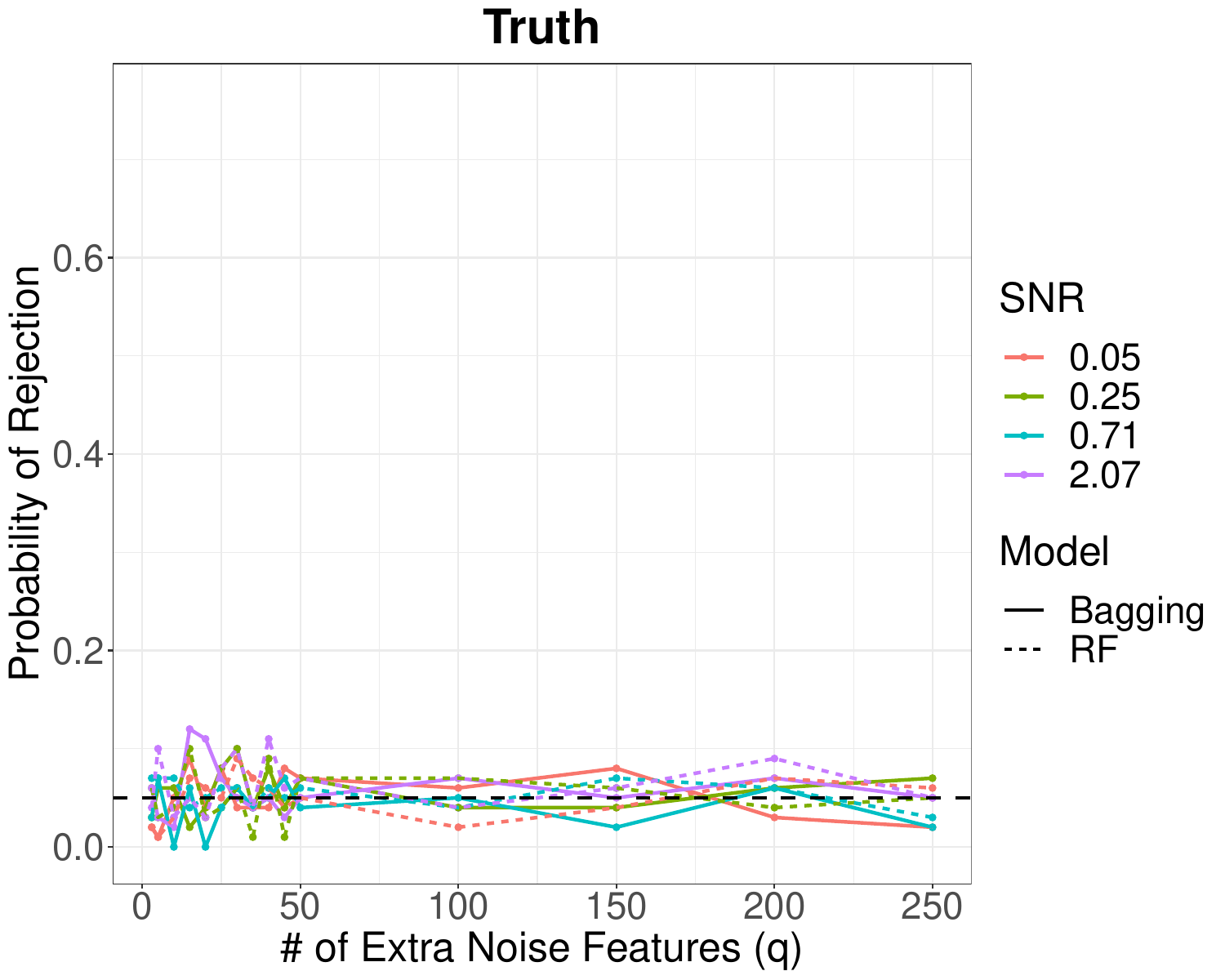}
\caption{Empirical rejection probabilities when testing the null hypothesis that $q$ additional noise features are predictively significant. Left: Replacements using random permutations; Middle: Replacements using MX knockoffs; Right: Replacements from the true conditional distribution. } \label{fig:HypTests}
\end{figure}

We consider 3 different types of randomized substitutes $\bm{x}_{2}^{\star}$. The first is a random permutation of the original $\bm{x}_{2}$, corresponding directly to the ``Permute-and-Relearn" approach. Since we know the true joint feature distribution in this simulation, random samples of $\bm{x}_{2}$ from the true conditional distribution can be obtained and used as the second type of random substitutes $\bm{x}_{2}^{\star}$.  On the other hand, in practice, this conditional distribution needs to be estimated.  Thus, as our third form of substitution we utilize the Model-X (MX) knockoff framework proposed by \cite{Candes2018} to generate $\bm{x}_{2}^{\star}$. In particular, we sample second-order multivariate Gaussian knockoff variables using the approximate semidefinite program construction \citep{Candes2018} as implemented in the \texttt{R} package \texttt{knockoff}. The entire procedure is repeated 100 times for each setting combination and each time we record whether the null hypothesis above was rejected.  Averaging across the 100 trials gives the proportion of times we reject $H_0$ and conclude that the $q$ additional noise features are significantly important.  Results are shown in Figure \ref{fig:HypTests}.

Importantly, note that because the $q$ additional noise features are conditionally independent of the response, intuitively, we should expect this test to reject only about $5\%$ of the time when $\alpha=0.05$, as it is here.  It is immediately obvious from the left plot in Figure \ref{fig:HypTests}, however, that in many settings, especially as increasingly many noise features are added, the rejection rates of the ``Permute-and-Relearn" approach lie well above that nominal level.  The heart of the issue here is that by permuting those $q$ features in $\bm{x}_2$, we are breaking not only their relationship to the response, but also their relationship to (i.e.\ correlation with) other features.  Also note, importantly, that this is not an issue arising from this particular test or set of hypotheses -- any test of conditional independence evaluated by comparing model performance when a subset of features are dropped or permuted will suffer the same issue. On the other hand, with random substitutes drawn from the true conditional distribution (right-most plot in Figure \ref{fig:HypTests}), the false rejection rates lie very near the pre-specified nominal level of 0.05.  The middle plot in Figure \ref{fig:HypTests} suggests MX knockoffs are able to adequately account for this correlation structure as here too we see rejections happening about 5\% of the time.

Thus, in keeping with the overall theme of our work, this demonstrates that even though relearning after permuting solves many of the initial issues discussed above, relearning alone does not fully resolve the issues associated with permutation-based variable importance.  We also stress that the alternative importance measures discussed above that specifically try to generate randomized replacement feature copies that preserve the between-feature relationship do not appear to suffer from such testing difficulties.  For a much more extended discussion on this topic, we refer the reader to \cite{Mentch2020}.

\subsection{Real World Data: Bike Sharing}

\begin{figure}
\centering
\includegraphics[height=12cm]{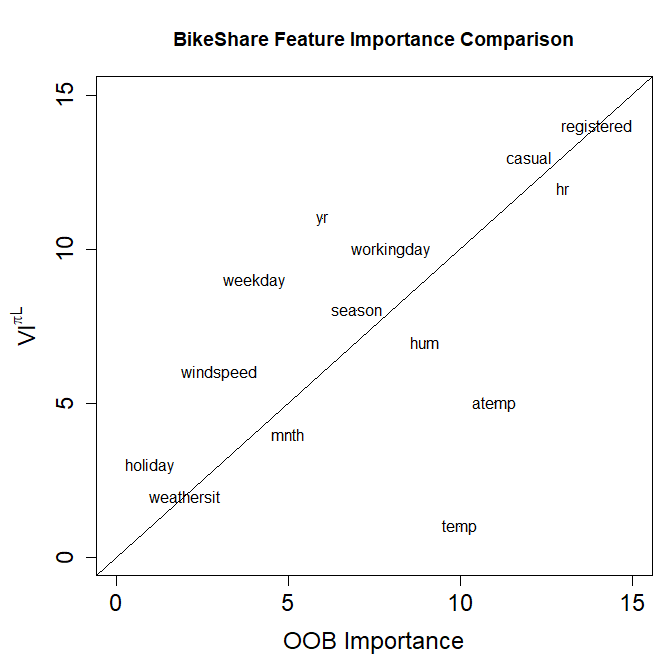}
\caption{A real-world comparison of OOB variable importance ranks (x-axis) with those obtained by permute-and-re-learn importance (y-axis) using random forests on the bikeshare data.} \label{fig:bike}
\end{figure}

To illustrate the real-world impact of the difference among these importance measures, Figure \ref{fig:bike} presents the importance rankings  of features in the hourly bike share data reported in \citet{Fanaee2013} as stored on the UCI repository. Here we predicted the log number of rentals each hour from 14 features using the \texttt{randomForest} package in \texttt{R}. We report the importance rank (ordered from least to most) of each feature as calculated by the default out-of-bag permutation measure, and as calculated by $\mbox{VI}^{\pi L}_j$ -- permuting the $j$th feature, but then learning a new model before assessing the change in accuracy. $\mbox{VI}^{\pi L}_j$ was chosen because it required only re-using the current learning method and maintained the dimension of the feature space.

Here we observe that while many of the features exhibit comparable ranks, there are some notable disagreements, \texttt{temp}, in particular, appearing important in OOB measures, but quite unimportant otherwise. A screening tool based on OOB importance might have selected this rather than \texttt{yr} to include in candidate features, possibly harming subsequent analysis.

\section{Partial Dependence Alternatives} \label{sec:GAMS}

While restricting to the data distribution creates a number of ways to redesign feature importance measures as illustrated in Section \ref{sec:alt}, this is less straightforward for plots of effects.  We now discuss briefly some alternatives that avoid extrapolation.

ICE plots can be restricted to a range of values consistent with the data distribution. In Figure \ref{fig:PD}, we have indicated regions with high data density by making the ICE plot lines solid over this portion of $x_1$. This device provides local information, but also serves to demonstrate the extent of dependence of $x_1$ on other covariates. However, we note here that we need some means of determining an appropriate range of values for each $x_1$. Originally, \cite{Goldstein2015} suggested labeling all observations on ICE plots, although this does not indicate the range over which the ICE plot is based on values near observed data.  In Figure \ref{fig:PD}, we have obtained this from 2-standard deviations within the Gaussian copula used to generate the data. However, lacking this information, obtaining these bounds requires us to estimate a conditional distribution.

Figure \ref{fig:ICEExtrap} suggests modifying partial dependence plots to average only the solid parts of each ICE line. This could be strongly misleading -- attributing changes due to $x_2$ rather to $x_1$.  \citet{Hooker2007} provides a re-interpretation of permutation diagnostics in terms of the functional ANOVA decomposition \eqref{eq:fanova}. In this framework, equivalents of partial dependence functions can be obtained by finding functions $f_1(x_1)$ and $f_{-1}(\bm{x}_{-1})$ to minimize
\[
\int \left(f(\bm{x}) - f_1(x_1) - f_{-1}(\bm{x}_{-1})\right)^2 p(\bm{x}) d\bm{x}
\]
in which $p(\bm{x})$ approximates the feature distribution and $f_{-1}(\bm{x}_{-1})$ is an unknown function of all features except $x_1$. \citet{Hooker2007} minimized a quasi-Monte Carlo approximation to this integral, but required an estimate of $p(\bm{x})$, which will likely not be accurate in high dimensions.  \citet{Chastaing2012} modified the estimation technique to include smoothing methods.  Similar structures were used in \citet{Mentch2017} to test for feature interactions.

A key problem for these methods is the need to jointly estimate $f_1(x_1)$ and the high dimensional $f_{-1}(\bm{x}_{-1})$, requiring specialized learning methods.  \citet{Lou2013}  and \citet{tan2018distill} instead fit a generalized additive model \citep{Wood2006} that includes only one-dimensional functions in the right hand side of \eqref{eq:fanova}. These can be fit to the data directly or may be used to examine the values of $f(\bm{x})$ as a model distillation method.  In Figure \ref{fig:PD}, our linear model fits explicitly fall within this class and provide a coherent, extrapolation-free, representation of the underlying function.

\section{Conclusions}




While na\"{i}ve permutation-based methods can be appealing, we have shown with some quite simple examples that they can give misleading results. We also identify the extrapolation behavior by flexible models as a significant source of error in these diagnostics. The precise biases that permute-and-predict methods produce will depend on the learning method employed as well as the specifics of the dependence of the features and the response function; see \citet{benard2021mda} for approaches to this analysis.

Alternative methods for summaries of feature importance all require further computational effort. $\mbox{VI}^{\pi L}_j$ still employs permutations, but learns a new model after permuting the features; \citet{Mentch2016} found that by maintaining the feature dimension, this made for a better statistical comparison than dropping a feature and re-learning as we have done in $\mbox{VI}^{D}_j$.  As demonstrated in earlier sections, an even better approach is to condition-and-relearn so that the randomized features used to replace those under investigation maintain at least approximate dependencies between the remaining features.  Recent work by \cite{berrett2020conditional} presents something of a hybrid approach between these that still utilizes permuted replacement features, but where those permutations are sampled in a non-uniform fashion informed by the conditional distribution.  Methods that avoid re-learning the response can be based on conditional permutation or simulation, but in general that still requires a model for $x_{ij}|\bm{x}_{i,-j}$; see \citet{Liu2018} for recent developments. Alternative measures of importance include generalizations of Sobol indices \citep{Hooker2007} and Shapley values \citep{Owen2014,lundberg2017unified}, although the calculation of these, if not undertaken carefully, can be subject to the same extrapolation bias that we identify here.

Beyond assigning importance scores, the use of permute-and-predict methods in PD and ICE plots are concerning for the same reasons. Alternatives to these are more challenging. Local explanation methods such as LIME and saliency maps avoid extrapolation, but do not provide a global representation of the learned model over the whole range of feature values; a threshhold function being a case in point. We also note that the counterfactual explanation methods explored in \citet{Wachter2017} may pose similar extrapolation problems, but have not explored this here.

 As an alternative visual diagnostic, the additive models explored in \citet{tan2018distill} produce effective graphical displays. \citet{Tan2018} found that these better represented the over-all behavior of the function than methods based on combining local explanations. However, specialized methods are still required to employ the diagnostics suggested in \citet{Hooker2007}.

\vspace{0.2cm}

{\bf Acknowledgments} This work was partially supported by NSF grants DMS-1712554, DMS-1712041 and DMS-2015400 as well as the Center for Research Computing at the University of Pittsburgh.

\bibliographystyle{chicago}
\bibliography{database}

\appendix

\section{Proofs of Results}

{\bf Proof of Theorem \ref{thm:lin}:}  We observe $E_{\pi} x_{ij}^{\pi,j} = \bar{x}_j$ and write
\[
\mbox{VI}^{\pi}_j = \sum_{i=1}^{N} (f(\bm{x}_i) - f(\bm{x}_i^{\pi,j}))^2 + 2 (y_i - f(\bm{x}_i))(f(\bm{x}_i) - f(\bm{x}_i^{\pi,j})).
\]
Examining the second term, when $f(\bm{x}_i) = \bm{x}_i \beta$ standard results give that $\sum_{i=1}^{N} (y_i - \bm{x}_i \hat{\beta})x_i = 0$ further $E_{\pi} \sum_{i=1}^N (y_i - \bm{x}_i \hat{\beta})x_{ij}^{\pi,j} = \sum_{i=1}^{N} (y_i - \bm{x}_i \hat{\beta}) \bar{x}_j = 0$.
Adding and subtracting $\bar{x}_j$, the first term can be simplified to
\[
E_{\pi} \hat{\beta}_j^2 \sum_{i=1}^{N} (x_{ij} - x_{ij}^{\pi,j})^2 = 2 \hat{\beta}_j^2 \sum_{i=1}^{N} (x_{ij} - \bar{x}_j)^2
\]
yielding the first result.  The remaining identities are the result of direct calculation. \\

\noindent {\bf Proof of Theorem \ref{thm:oracle}:} We begin by observing that
\begin{align*}
\argmin_{m} E_{y|\bm{x}^*} (y - m)^2 & = \argmin_{m} E_{y|\bm{x}_{-j}} (y - m)^2 \\
& = E_{y|\bm{x}_{-j}} y \\
& = E_{x_j | \bm{x}_{-j}} E_{y|\bm{x}} y \\
& = f_{-j}(\bm{x}_{-j}).
\end{align*}
This is the quantity we estimate directly in $\mbox{VI}^C_j$. Observerving that $x_j^{\pi,j}$, $x_j^{c,j}$ and, in the case of $\mbox{VI}^D_j$ replacing $x_j$ with any constant, all satisfy $x_j^* \perp y|\bm{x}_{-j}$, yields the result.

\noindent {\bf Proof of Theorem \ref{thm:drop.equal}:}  Observe that for any invertible matrix $A$, regressing on $XA$ is equivalent to regressing on $X$. 

For simplicity, we will assume that the feature matrix $X$ includes a column of 1's to allow for an intercept. Writing $X \hat{\bm{\beta}} = X^{-j} (\hat{\bm{\beta}}^{-j} + \hat{\gamma}^{j} \hat{\beta}_j) + \beta_j \bm{\delta}_j$ we have
\begin{align*}
f^{-j}(\bm{x}) & = \bm{x}(X^{-jT}X^{-j})^{-1} X^{-jT}(X^{-j} (\hat{\bm{\beta}}^{-j} + \hat{\gamma}^{j} \hat{\beta}_j) + \beta_j \bm{\delta}_j + \bm{e}) \\ & = \bm{x}^{-j} (\hat{\bm{\beta}}^{-j} + \hat{\gamma}^{j} \hat{\beta}_j)
\end{align*}
where $\bm{e} = \bm{y} -  X \hat{\bm{\beta}}$ are orthogonal to $X^{-j}$, as is $\bm{\delta}_j$.
Then writing
\[
\mbox{VI}^D_j = \sum_{i=1}^N (\bm{x}_i(\hat{\bm{\beta}} - \hat{\bm{\beta}}^{-j}))^2 + 2 (y_i - \bm{x}_i\hat{\bm{\beta}})\bm{x}_i(\hat{\bm{\beta}} - \hat{\bm{\beta}}^{-j})
\]
we see the second term is zero by standard results and we obtain the result by substitution.

The result for $\mbox{VI}^{\pi L}_j$ can be shown by first centering the $x_{ij}$ after which $E_{\pi} \sum_{i=1}^{N} x_{i j}^{\pi,j} y_i = E_{\pi} \sum_{i=1}^{N} x_{i j}^{\pi j} x_{it k} = 0$. Hence, $E_{\pi} f^{\pi,j}(\bm{x}) = f^{-j}(\bm{x}) + o_p(1/N)$ and the result follows from that for $\mbox{VI}^D_j$.

The result for $\mbox{VI}^{C L}_j$ is obtained by the same calculation, after subtracting $\gamma_{0j} + \bm{x}_{i,-j} \bm{\gamma}_j$ from $x_{ij}^{c,j}$.

The result for $\mbox{VI}^C_j$ follows from the calculations  in the proof of Theorem \ref{thm:lin} after observing that $ E( x_{ij} - x_{ij}^{c,j} )^2 = 2 \mbox{var}(x_{ij}|\bm{x}_{-j})$.

\end{document}